\documentclass[preprint,12pt]{elsarticle}
\journal{Computer Physics Communications}

\usepackage[colorlinks, urlcolor=blue,linkcolor=blue, anchorcolor=blue, citecolor=blue]{hyperref}
\hypersetup{bookmarksopen=true}

\usepackage[title,titletoc]{appendix}
\usepackage{amssymb}
\usepackage{amsmath}
\usepackage{graphicx}
\usepackage{subfigure}
\usepackage{multirow}
\usepackage{color}
\usepackage{appendix}
\usepackage{fancyhdr}
\usepackage[utf8]{inputenc}
\usepackage{caption}
\DeclareCaptionFont{white}{\color{white}}
\DeclareCaptionFormat{listing}{\colorbox{gray}{\parbox{\textwidth}{#1#2#3}}}
\biboptions{sort&compress}

\graphicspath{{fig/}}

\newcounter{bla}

\newcommand{\V}{n}
\newcommand{\n}{{\cal N}}
\newcommand{\ed}{{\cal E}}
\newcommand{\p}{{\cal P}}
\newcommand{\peq}{{\cal P}_{0}}
\newcommand{\s}{{\cal S}}
\newcommand{\code}{{\sc BEShydro}}
\newcommand{\tauR}{\tau_\mathrm{rel}}

\begin{document}
\begin{frontmatter}

\title{(3+1)-dimensional dissipative relativistic fluid dynamics at non-zero net baryon density}

\author[a]{Lipei Du\corref{author}}
\author[a,b]{Ulrich Heinz\corref{email}}
\cortext[author] {Email: du.458@osu.edu (corresponding author)}
\cortext[email] {Email: heinz.9@osu.edu}
\address[a]{Department of Physics, The Ohio State University, Columbus, OH 43210-1117, USA}
\address[b]{Institut f\"ur Theoretische Physik, J.~W.~Goethe Universit\"at, Max-von-Laue-Str. 1, D-60438 Frankfurt am Main, Germany}
\date{\today}

\begin{abstract}
    Heavy-ion collisions at center-of-mass energies between 1 and 100~GeV/nucleon are essential to understand the phase diagram of QCD and search for its critical point. At these energies the net baryon density of the system can be high, and simulating its evolution becomes an indispensable part of theoretical modeling. We here present the (3+1)-dimensional diffusive relativistic hydrodynamic code \code\ which solves the equations of motion of second-order Denicol-Niemi-Molnar-Rischke (DNMR) theory, including bulk and shear viscous currents and baryon diffusion currents. {\sc BEShydro} features a modular structure that allows to easily turn on and off baryon evolution and different dissipative effects and thus to study their physical effects on the dynamical evolution individually. An extensive set of test protocols for the code, including several novel tests of the precision of baryon transport that can also be used to test other such codes, is documented here and supplied as a permanent part of the code package.
\end{abstract}

\end{frontmatter}

\newpage

\noindent
{\bf PROGRAM SUMMARY}
\vspace{0.5cm}\\
\begin{small}
\noindent
{\em Manuscript Title:} (3+1)-dimensional dissipative relativistic fluid dynamics at non-zero net baryon density \\
{\em Authors:} Lipei Du, Ulrich Heinz \\
{\em Program Title:} \code \\
{\em Licensing provisions:} GPLv3 \\
{\em Programming language:} C++  \\
{\em Computer:} Laptop, desktop, cluster  \\
{\em Operating system:} GNU/Linux distributions, Mac OS X \\
{\em Memory usage:} For a $121\times121\times121$ grid, 0.57 GB (conserved variables), 0.003 GB (EOS4 tables), 0.035 GB (dynamical sources)
{\em Classification:} 12 Gases and Fluids, 17 Nuclear Physics\\
{\em External routines/libraries:} GNU Scientific Library (GSL)\\
{\em Nature of problem:} (3+1)-dimensional dynamical evolution of hot and dense matter created in relativistic heavy-ion collisions using second-order dissipative relativistic fluid dynamics, including evolution of net baryon number and its dissipative diffusion current. \\
{\em Solution method:} Runge-Kutta Kurganov-Tadmor algorithm\\
{\em Running time:} A test run with baryon diffusion on a 2-dimensional grid $121\times121$ uses 0.32 sec/time step, and on a 3-dimensional grid $121\times121\times121$ uses 46.08 sec/time step on a MacBook Pro with a 2.7 GHz Intel Core i5 processor and 8 GB 1867 MHz DDR3 memory.\\
\end{small}

\newpage
\tableofcontents
\newpage
\section{Introduction}

 Over the last decade, second-order dissipative relativistic fluid dynamics (RFD) \cite{ISRAEL1976310, rspa.1977.0155, Israel:1979wp} has developed into a powerful and phenomenologically very successful tool for the description of the dynamical evolution of the hot and dense matter created in relativistic heavy-ion collisions \cite{Heinz:2005bw, Song:2008si, Song:2007fn, Romatschke:2007mq, PhysRevC.82.014903, Gale:2012rq, Martinez:2010sc, Florkowski:2010cf, Martinez:2012tu, Florkowski:2014bba, Karpenko:2013wva, Molnar:2009tx, PhysRevD.85.114047}. It is an effective theory for the macroscopic evolution of the conserved quantities of the system (energy, momentum, baryon number, strangeness, isospin and electric charge), coupled to dissipative flows whose dynamics is controlled by the competition between microscopic scattering (which moves the system closer to local thermodynamic equilibrium) and macroscopic expansion (which tends to drive the system away from local equilibrium). In heavy-ion collisions, dissipative fluid dynamics describes well the evolution of the quark-gluon plasma (QGP), a strongly coupled plasma that exhibits almost perfect liquid behavior \cite{Kolb:2000fha, Huovinen:2001cy, Song:2010mg, Romatschke:2007mq, Dusling:2007gi, Luzum:2008cw} but exists only at temperatures above about 150~MeV. To obtain quantitatively precise predictions for heavy-ion collisions, the fluid dynamic stage must be initialized with the output from some microscopic model for the pre-hydrodynamic stage at very early times, when the energy deposited in the collision fireball is still so far away from local momentum isotropy and thermal equilibrium that a hydrodynamic language can not yet be applied \cite{Liu:2015nwa, Kurkela:2018wud, Kurkela:2018vqr, vanderSchee:2013pia, Shen:2017bsr, Du:2018mpf}, and it must be coupled to a microscopic kinetic description of the hadronic rescattering stage at the end of the fireball evolution \cite{Bass:1998ca, Bleicher:1999xi, Weil:2016zrk, Novak:2013bqa, PhysRevC.61.024901}, after the QGP has converted back into a gas of hadrons and hadronic resonances which quickly moves towards ``kinetic freeze-out'' when all strong interactions cease and the energies and momenta of all strongly interacting particles stop changing.

The initial development of dissipative RFD ignored the evolution of conserved currents such as net baryon number and strangeness because the community's attention was focused on experiments performed at the highest available collision energies at the Relativistic Heavy Ion Collider (RHIC) at Brookhaven National Laboratory and the Large Hadron Collider (LHC) at CERN, at which the colliding atomic nuclei are largely transparent to each other, creating a system of approximately zero net baryon number, strangeness, and isospin charge near midrapidity in the center of mass frame (formed by the interaction of low-$x$ gluons from the colliding nuclei which carry no conserved charges) whereas the net baryon number and electric charges brought into the collision by the valence quarks within the incoming nuclei cannot be stopped and end up mostly at far forward and backward rapidities where they are very hard to measure experimentally. Only with the Beam Energy Scan (BES) program at RHIC \cite{Aggarwal:2010cw}, in which heavy-ion collisions were studied at lower collision energies where some of the incoming baryon charge gets stopped near midrapidity, became the need urgent for including the dynamics of the baryon number and other conserved charge currents in the hydrodynamic description. These developments are also relevant for the theoretical description of future experiments at NICA \cite{Sissakian:2009zza} and FAIR \cite{SPILLER2006305,CHATTOPADHYAY2014267}.

We here describe a new (3+1)-dimensional dissipative RFD code, which we call \code, that includes the evolution of the net baryon charge and diffusion currents.\footnote{%
    The code is open source and can be freely downloaded from \url{https://github.com/LipeiDu/BEShydro}}
\code\ evolved out of the CPU version of the code {\sc GPU-VH} \cite{Bazow:2016yra}, by adding a number of additional dissipative terms \cite{PhysRevD.85.114047, PhysRevLett.105.162501, Molnar:2009tx} describing the evolution of the bulk and shear viscous pressures, in addition to the evolution equations for the net baryon charge and diffusion currents. As far as we know, at this point in time the only other code that shares all of the main features of \code\ is the latest version of {\sc MUSIC} \cite{Denicol:2018wdp}, while other codes (e.g. \cite{Karpenko:2013wva, Bazow:2016yra, Pang:2018zzo}) so far ignore the evolution of the net baryon diffusion current or, in some cases, even that of the net baryon charge. \code\ has been developed completely independently of {\sc MUSIC}; it can thus serve as a platform for detailed code validations and comparisons, even if in future applications the two codes will likely be applied to different collision systems, using different initialization modules and hadronic afterburners.

The purpose of this document is to review the physics ingredients needed for describing the hydrodynamic evolution of systems with non-zero conserved charges (such as baryon number) and to provide a detailed description of the structure and performance of \code.
We start from the most general form of the equations solved by \code\ in Sec.~\ref{phys-framework}, describe the numerical scheme used by the code in  Sec.~\ref{numericalscheme}, and include a number of validation tests in Sec.~\ref{numericaltests}, using specific, highly symmetric settings in which analytic or semi-analytic solutions of the hydrodynamic evolution equations have been found. Some of the novel tests of the charge transport sector of the code described here should be useful for the developers of other codes with the ability to describe dissipative effects related to charge diffusion. At this point \code\ evolves only a single conserved charge (net baryon number); future generalization to include strangeness and isospin \cite{Greif:2017byw} is expected to be conceptually straightforward.

\section{Relativistic hydrodynamics}\label{phys-framework}
%
%
In this section the physics ingredients of \code\ are described: the propagated physical quantities, their equations of motion, and several variants of the Equation of State (EoS) controlling the expansion of the liquid formed in relativistic heavy-ion collisions.

\subsection{Equations of motion}

\subsubsection{Conservation laws}

Hydrodynamics is a macroscopic theory describing the space-time evolution of the 14 components of the energy-momentum tensor $T^{\mu\nu}(x)$ and the net (baryon) charge current $N^\mu(x)$.\footnote{%
     For $N_C>1$ conserved charges, the number of evolved quantities increases to $10{\,+\,}4N_C$.}
Five evolution equations arise from the conservation laws for energy, momentum, and the baryon charge \cite{Molnar:2009tx}:
\begin{eqnarray}
d_{\mu}T^{\mu\nu} &\equiv&\frac{1}{\sqrt{g}}\partial_{\mu}(\sqrt{g}T^{\mu\nu})+\Gamma^{\nu}_{\mu\lambda}T^{\mu\lambda}=0\;,
\label{hydro_eqs_T}\\
d_{\mu}N^{\mu} &\equiv&\frac{1}{\sqrt{g}}\partial_{\mu}(\sqrt{g}N^{\mu})=0\;.
\label{hydro_eqs_n}
\end{eqnarray}
Here $d_\mu$ ($\mu=0,1,2,3$) stands for the covariant derivative in a general system of space-time coordinates, with metric tensor $g^{\mu\nu}$ defined with negative signature (``mostly minus'' convention $(+,-,-,-)$), $g\equiv-\det{(g_{\mu\nu})}$, and the Christoffel symbols (see, e.g., \cite{carroll2004spacetime})
\begin{equation}
    \Gamma^{\mu}_{\alpha\beta}\equiv \frac{1}{2} g^{\mu\nu} \bigl(\partial_{\beta}g_{\alpha\nu}
         +\partial_{\alpha}g_{\nu\beta}
         -\partial_{\nu}g_{\alpha\beta}\bigr)
    =\Gamma^{\mu}_{\beta\alpha}\;.
\end{equation}

The 14 independent components of $T^{\mu\nu}$ and $N^\mu$ are more physically defined in terms of the hydrodynamic decomposition of these tensors \cite{landau2013fluid},
\begin{eqnarray}
    T^{\mu\nu} &=& \ed u^{\mu}u^{\nu}-(\peq+\Pi)\Delta^{\mu\nu}+\pi^{\mu\nu}\;,\label{eq-dec-T}\\
    N^{\mu} &=& \n u^{\mu}+\V^{\mu}\;. \label{eq-dec-N}
\end{eqnarray}
Here the flow 4-velocity $u^\mu(x)$, with $u^\mu u_\mu=1$, is defined as the time-like eigenvector of the energy-momentum tensor,
\begin{equation}
    T^{\mu\nu}u_\nu=\ed u^\mu\;,
\end{equation}
and specifies the local rest frame (LRF) of the fluid at point $x$ (the so-called ``Landau frame''). The tensors $u^{\mu}u^{\nu}$ and $\Delta^{\mu\nu} \equiv g^{\mu\nu} - u^{\mu}u^{\nu}$ then are projectors on the temporal and spatial directions in the LRF. $\ed$ and $\n$ are the energy and net baryon density in the LRF which can be obtained as the following projections of $T^{\mu\nu}$ and $N^\mu$:
\begin{equation}
    \ed = u_\mu T^{\mu\nu} u_\nu\;,\qquad \n = u_\mu N^{\mu}\;.\label{eq-landau}
\end{equation}
From these, the local equilibrium pressure $\peq$ is obtained through the EoS $\peq=\peq(\ed,\n)$. The shear stress $\pi^{\mu\nu}$, the bulk viscous pressure $\Pi$, and the baryon diffusion current $n^\mu$ are dissipative flows describing deviations from local equilibrium.

Using the decomposition (\ref{eq-dec-T},\ref{eq-dec-N}) the conservation laws (\ref{hydro_eqs_T},\ref{hydro_eqs_n}) can be brought into the physically intuitive form \cite{Jeon:2015dfa} 
\begin{eqnarray}
 D\n &=& -\n\theta -\nabla_\mu n^\mu\;,
\label{eq:vhydro-N}
\\
 D\ed &=& -(\ed{+}\peq{+}\Pi)\theta + \pi_{\mu\nu}\sigma^{\mu\nu}\;,
\label{eq:vhydro-E}
\\
  (\ed{+}\peq{+}\Pi)\, Du^\mu &=& \nabla^\mu(\peq{+}\Pi)
    - \Delta^{\mu\nu} \nabla^\sigma\pi_{\nu\sigma} + \pi^{\mu\nu} Du_\nu\;.
\label{eq:vhydro-u}
\end{eqnarray}
Here $D=u_\mu d^\mu$ denotes the time derivative in the LRF, $\theta=d_\mu u^\mu$ is the scalar expansion rate, $\nabla^\mu=\partial^{\langle\mu\rangle}$ (where generally $A^{\langle\mu\rangle} \equiv \Delta^{\mu\nu}A_\nu$) denotes the spatial gradient in the LRF, and $\sigma^{\mu\nu} = \nabla^{\langle\mu}u^{\nu\rangle}$ (where generally $B^{\langle\mu\nu\rangle} \equiv \Delta^{\mu\nu}_{\alpha\beta} B^{\alpha\beta}$, with the traceless spatial projector $\Delta^{\mu\nu}_{\alpha\beta} \equiv \frac{1}{2} (\Delta^\mu_\alpha \Delta^\nu_\beta + \Delta^\nu_\alpha \Delta^\mu_\beta) - \frac{1}{3} \Delta^{\mu\nu} \Delta_{\alpha\beta}$) is the shear flow tensor. While these equations clearly exhibit the physics in the LRF (which varies from point to point), \code\ solves the conservation laws (\ref{hydro_eqs_T},\ref{hydro_eqs_n}) in a fixed global computational frame. Their explicit form in the global frame is discussed in Sec.~\ref{sec2.1.3}.

The 5 conservation laws (\ref{eq:vhydro-N}-\ref{eq:vhydro-u}) are sufficient to determine the energy and baryon density, $\ed$ and $\n$, together with the 3 independent components of the flow velocity $u^\mu$, as long as the shear stress $\pi^{\mu\nu}$, the bulk viscous pressure $\Pi$, and the baryon diffusion current $n^\mu$ vanish.\footnote{\label{fn2}%
     Note that the shear stress is traceless, $\pi^\mu_\mu=0$, and both $\pi^{\mu\nu}$ and $n^\mu$ have only spatial components in the LRF, $u_\mu \pi^{\mu\nu}=\pi^{\mu\nu} u_\nu = u_\mu n^\mu =0$. $\pi^{\mu\nu}$, $\Pi$, and $n^\mu$ thus describe 5+1+3=9 dissipative degrees of freedom.}
Their evolution is not directly constrained by conservation laws but controlled by the competition between microscopic scattering processes (which drive the system towards local equilibrium and the dissipative flows to zero) and the macroscopic expansion (which drives the system away from equilibrium and the dissipative flows away from zero). Their evolution is thus controlled by both micro- and macroscopic physics. One way to obtain their evolution equations is DNMR theory \cite{PhysRevD.85.114047, PhysRevLett.105.162501, Molnar:2009tx} which uses the method of moments of the Boltzmann equation and which we employ here. 

Expressed through its natural variables, i.e.\ the temperature $T$ and baryon chemical potential $\mu$, the equilibrium pressure $\peq(\ed,\n)=\peq(T,\mu)$ is recognized as the grand-canonical thermodynamic potential for a system with temperature $T(\ed,\n)$ and chemical potential $\mu(\ed,\n)$. In principle, $T$ and $\mu$ are not needed for the hydrodynamic evolution, but they may be required to compute certain signatures of the evolving fluid (such as the spectrum of electromagnetic radiation emitted during its evolution or the spectrum of hadrons into which it decays at the end of the life of the quark-gluon plasma phase), and in \code\ the driving force for net baryon number diffusion is formulated in terms of the gradient of $\mu/T$ rather than that of the net baryon density. Also, the transport coefficients controlling the evolution of the dissipative flows are most naturally expressed as functions of $T$ and $\mu$ since they are defined as response functions of the thermal equilibrium system described by the potential $\peq(T,\mu)$. Different versions of the EoS $\peq(T,\mu)$ or $\peq(\ed,\n)$ used in \code\ will be described in Sec.~\ref{subsec-eos}.

\subsubsection{Evolution equations for the dissipative flows}
\label{sec2.1.2}

In \code\ the dissipative flows are evolved with DNMR theory \cite{Molnar:2009tx, PhysRevLett.105.162501, PhysRevD.85.114047}. While the equations of motion in this theory are derived from the Boltzmann equation which is applicable only to weakly coupled systems \cite{Arnold:2002zm}, the hydrodynamic description is an effective theory which is generic and applicable also in the strong coupling regime where the Boltzmann equation is not valid \cite{Baier:2007ix}. When applying the DNMR equations to the fluid produced in nuclear collisions, which appears to be strongly coupled, one must replace its material properties, i.e. the EoS and transport coefficients, by those for real QCD matter.

In the framework of DNMR theory, the dissipative transport equations are given by the following relaxation equations:
\begin{eqnarray}
    \tau_{\Pi }D{\Pi}+\Pi &=& \Pi_{\mathrm{NS}} +\mathcal{J}+\mathcal{K}+\mathcal{R}\;,\label{eq-relax-Pi}\\
    \tau_{n}(Dn)^{\left\langle \mu \right\rangle}+n^{\mu} &=& n^{\mu}_{\mathrm{NS}}+\mathcal{J}^{\mu }+\mathcal{K}^{\mu }+\mathcal{R}^{\mu }\;,\label{eq-relax-n}\\
    \tau _{\pi }(D\pi)^{\left\langle \mu \nu \right\rangle }+\pi ^{\mu \nu }
    &=& \pi^{\mu \nu }_{\mathrm{NS}}+\mathcal{J}^{\mu \nu }+\mathcal{K}^{\mu \nu }+\mathcal{R}^{\mu \nu }\;.\label{eq-relax-pi}
\end{eqnarray}
Here $(Dn)^{\langle\mu\rangle}\equiv \Delta^{\mu\nu} Dn_\nu$ and $(D\pi)^{\langle\mu\nu\rangle} \equiv \Delta^{\mu\nu}_{\alpha\beta} D\pi^{\alpha\beta}$, ensuring that all terms are purely spatial in the LRF and, where applicable, traceless. $\tau _{\Pi }$, $\tau_{n}$, and $\tau _{\pi}$ are the relaxation times for $\Pi$, $\V^\mu$, and $\pi^{\mu\nu}$, respectively. They control how fast the dissipative flows relax to their Navier-Stokes limits \cite{Jeon:2015dfa}:
\begin{eqnarray}
    \Pi_{\mathrm{NS}} &=& -\zeta \theta\;,\label{eq-ns-Pi}\\
    n^{\mu}_{\mathrm{NS}} &=& \kappa_n \nabla^{\mu}\left(\frac{\mu}{T}\right)\;,\label{eq-ns-n}\\
    \pi ^{\mu \nu }_{\mathrm{NS}} &=& 2\eta \sigma ^{\mu \nu }\;,\label{eq-ns-pi}
\end{eqnarray}
where $\zeta$, $\kappa_n$, and $\eta$ are the bulk viscosity, baryon diffusion coefficient, and shear viscosity, respectively, describing the first-order response of the dissipative flows to their driving forces, the (negative of the) scalar expansion rate $\theta$, the spatial gradient of $\mu/T$ in the LRF, $\nabla^{\mu}(\mu/T)$, and the shear flow tensor  $\sigma ^{\mu\nu}$, respectively, which drive the system away from local equilibrium. 

The scalar, vector and tensor source terms on the r.h.s. of Eqs.~(\ref{eq-relax-Pi}-\ref{eq-relax-pi}), $\mathcal{J},\; \mathcal{K},\; \mathcal{R},\; \mathcal{J}^{\mu},\; \mathcal{K}^{\mu},\; \mathcal{R}^{\mu},\; \mathcal{J}^{\mu \nu },\; \mathcal{K}^{\mu \nu }$ and $\mathcal{R}^{\mu \nu }$ contain terms of second order in the small parameters Knudsen number (ratio between a characteristic microscopic and macroscopic time or length scale of the fluid) and inverse Reynolds number (ratio between dissipative quantities and local equilibrium values). According to the notation established in \cite{PhysRevD.85.114047}, the Navier-Stokes and $\mathcal{K}$ terms on the r.h.s. of Eqs.~(\ref{eq-relax-Pi}-\ref{eq-relax-pi}) are of first and second order in the Knudsen number(s), respectively, the  $\mathcal{J}$ terms are of order Knudsen number times inverse Reynolds number, and the $\mathcal{R}$ terms are of second order in the inverse Reynolds number(s). Their explicit expressions can be found in Ref.~\cite{PhysRevD.85.114047}. Following the arguments in Ref.~\cite{Bazow:2016yra}, we here include only a subset of the $\mathcal{J}$ terms. As \code\ is ultimately designed for precision studies of relativistic heavy-ion collisions, one should perhaps not put too much blind trust into these arguments and rather check their validity; on the other hand, adding the missing second-order source terms to the code at a later time should be straightforward (even if additional code stability tests may be needed). Future code updates will include additional terms as required by specific applications.

As implemented in the code, the thus simplified relaxation equations for the dissipative flows read
\begin{eqnarray}
    \tau_{\Pi}D\Pi+\Pi &=&
        -\zeta\theta -\delta_{\Pi\Pi}\Pi\theta
        +\lambda_{\Pi\pi}\pi^{\mu\nu}\sigma_{\mu\nu}\;,
\label{eq-Pi-simple}
\\
    \tau _{n} D n^\mu+n^{\mu } &=&\text{ } \kappa_n \nabla^{\mu }\left(\frac{\mu}{T}\right) -\tau_n n_{\nu }\omega ^{\nu \mu }-\delta_{nn}n^{\mu }\theta \nonumber\\
      &-&\lambda _{nn}n_{\nu }\sigma ^{\mu \nu } - \tau_n n^\nu u^\mu D u_\nu\;,
\label{eq-n-simple}
\\
    \tau _{\pi }D\pi^{\mu \nu}+\pi ^{\mu \nu }&
        =&2\eta \sigma ^{\mu \nu }
        +2\tau_{\pi}\pi^{\langle\mu}_{\lambda}\omega^{\nu\rangle\lambda}
        -\delta_{\pi\pi}\pi^{\mu\nu}\theta\nonumber\\
        &-&\tau_{\pi\pi}\pi^{\lambda\langle\mu}\sigma^{\nu\rangle}_{\lambda}
        +\lambda_{\pi\Pi}\Pi\sigma^{\mu\nu}
        \nonumber \\
        &-&\tau_{\pi}( \pi ^{\lambda \mu }u^{\nu }+\pi ^{\lambda \nu}u^{\mu }) Du_{\lambda }\;; 
\label{eq-pi-simple}
\end{eqnarray}
here $\omega^{\mu\nu}=\frac{1}{2}(\nabla^\mu u^\nu-\nabla^\nu u^\mu)$ is the vorticity tensor. The additional transport coefficients $\delta_{\Pi\Pi}$, $\lambda_{\Pi\pi}$, $\tau_n$ etc. will be discussed in Sec.~\ref{transcoeff}. In Eqs.~(\ref{eq-n-simple}) and (\ref{eq-pi-simple}) we removed the transversality constraints on the l.h.s. by using footnote \ref{fn2} and
\begin{equation}
(D{n})^{\left\langle \mu \right\rangle} = \Delta^\mu_\nu D n^\nu = D n^\mu + u^\mu n^\nu D u_\nu\;,
\end{equation}
as well as its analog for $\pi^{\mu\nu}$, moving the extra terms as additional source terms to the r.h.s.

It is worth pointing out that in Eqs.~(\ref{eq-Pi-simple})-(\ref{eq-pi-simple}) we have followed Ref.~\cite{Denicol:2018wdp} in ignoring terms describing the direct influence of baryon diffusion, $n^\mu$, on the evolution of the shear and bulk viscous stresses, $\pi^{\mu\nu}$ and $\Pi$. Baryon evolution still affects the evolution of the system indirectly through the EoS. In this approach it has been shown \cite{Denicol:2018wdp, Du:2018mpf} that, while dissipative baryon diffusion effects directly influence the net-proton distributions, its indirect effects on the distributions of mesons and charged hadrons are negligible. It might be interesting to study to which extent second-order couplings between baryon diffusion and viscous stresses can modify this conclusion.

\subsubsection{Evolution equations in Milne coordinates}
\label{sec2.1.3}

Up to this point the formalism is completely general. For application to ultra-relativistic heavy-ion collisions we need the specific form of the evolution equations in Milne coordinates $x^{\mu}=(\tau,x,y,\eta_s)$ which are best adapted to the relativistic collision kinematics and subsequent almost boost-invariant longitudinal flow pattern \cite{PhysRevD.27.140}. In terms of Cartesian coordinates $(t,x,y,z)$ the longitudinal proper time $\tau$ and space-time rapidity $\eta_s$ are defined as
\begin{equation}
    \tau=\sqrt{t^2-z^2}\;,\qquad \eta_{s}=\frac{1}{2}\ln{\left(\frac{t+z}{t-z}\right)}\;.
\end{equation}
The mid-rapidity point $z=\eta_s=0$ at $\tau=0$ defines the collision point in the global (computational) frame. In Milne coordinates the metric tensor is
\begin{equation}
    g^{\mu\nu} =\mathrm{diag}\bigl(1,-1,-1,-1/\tau^2\bigr)\;,
\end{equation}
the fluid four-velocity is $u^\mu =(u^\tau,u^x,u^y,u^\eta)$, and the four-derivative is $\partial_{\mu}=(\partial_\tau,\partial_x,\partial_y,\partial_\eta)$.\footnote{%
        In all sub- and superscripts $\eta$ is short for $\eta_s$.}
The metric has the following non-vanishing Christoffel symbols:
\begin{equation}
    \Gamma^{\eta}_{\tau\eta}=\Gamma^{\eta}_{\eta\tau}=\frac{1}{\tau}\;,\qquad \Gamma^{\tau}_{\eta\eta}=\tau\;.\label{eq-chiris}
\end{equation}
Plugging them into Eqs.~(\ref{hydro_eqs_T},\ref{hydro_eqs_n}) we obtain the conservation laws in Milne coordinates:
\begin{eqnarray}
{\partial }_{\mu }T^{\mu\tau}& =&-\frac{1}{\tau}(T^{\tau\tau}+\tau^{2}T^{\eta\eta})\, , \label{relEqs_Tmut}\\
{\partial }_{\mu }T^{\mu x}& =&-\frac{1}{\tau}T^{\tau x}\, , 
\\ 
{\partial }_{\mu }T^{\mu y}&=&-\frac{1}{\tau}T^{\tau y}\, , 
\\ 
{\partial }_{\mu }T^{\mu \eta }&=&-\frac{3}{\tau}T^{\tau \eta}\ ,
\label{dT0i1}
\\
\partial_\mu N^{\mu}&=&-\frac{1}{\tau} N^\tau\;.
\label{relEqs_n}
\end{eqnarray}

Introducing the convective time derivative $d \equiv u^\mu \partial_\mu$, the relaxation equations can be written as
\begin{eqnarray}
    d\Pi&=&-\frac{\zeta}{\tau_\Pi}\theta-\frac{\Pi}{\tau_\Pi}-I_\Pi \label{relEqs_Pi}
\;,\\
    d \V^\mu &=& \frac{\kappa_n}{\tau_\V} \nabla^{\mu} \left(\frac{\mu}{T}\right) - \frac{\V^{\mu}}{\tau_{\V}} - I_\V^\mu 
    - G_\V^\mu\label{relEqs_nmu}\,,
    \\
    d\pi^{\mu\nu}&=&\frac{2\eta}{\tau_{\pi}}\sigma^{\mu\nu}-\frac{\pi^{\mu\nu}}{\tau_{\pi}}-I^{\mu\nu}_{\pi}-G^{\mu\nu}_{\pi}\,,
    \label{relEqs_pi}
\end{eqnarray}
with the shorthand notations $G^{\mu\nu}_{\pi}\equiv u^{\alpha}\Gamma^{\mu}_{\alpha\beta} \pi^{\beta\nu} + u^{\alpha}\Gamma^{\nu}_{\alpha\beta}\pi^{\beta\mu}$ and $G^\mu_n = u^\alpha \Gamma^\mu_{\alpha\beta} n^\beta$ for the geometrical source terms obtained when splitting the covariant LRF time derivative $D$ in Eqs.~(\ref{eq-Pi-simple})-(\ref{eq-pi-simple}) into the convective time derivatives $d$ and a remainder (for example, $D n^\mu \equiv u^\alpha d_\alpha n^\mu = u^\alpha (\partial_\alpha n^\mu + \Gamma^\mu_{\alpha\beta} n^\beta)$). The $I$-terms are explicitly
\begin{eqnarray}
I_{\Pi}&\equiv& \frac{\delta_{\Pi\Pi}}{\tau_\Pi}\Pi\theta-\frac{\lambda_{\Pi\pi}}{\tau_\Pi}\pi^{\mu\nu}\sigma_{\mu\nu}\;,\label{IPiterms}
\\
I^\mu_n &\equiv& I_1^\mu+ \frac{\delta _{nn}}{\tau _{n}} I_2^\mu+ I_3^\mu+ \frac{\lambda _{nn}}{\tau _{n}} I_4^\mu\;, \label{Interms}\\
I^{\mu\nu}_{\pi}&\equiv& I^{\mu\nu}_{1}
+\frac{\delta_{\pi\pi}}{\tau_\pi}I^{\mu\nu}_{2}
-I^{\mu\nu}_{3}
+\frac{\tau_{\pi\pi}}{\tau_\pi}I^{\mu\nu}_{4}
-\frac{\lambda_{\pi\Pi}}{\tau_\pi}\Pi\sigma^{\mu\nu}\;,\label{Ipiterms}
\end{eqnarray}
with 
\begin{eqnarray}
\label{I1-I4-mu}
  &&\!\!\!\!\!\!
    I_1^\mu = u^\mu n^\nu Du_\nu,\quad 
    I_2^\mu = n^{\mu}\theta,\quad
    I_3^\mu = n_{\nu}\omega^{\nu\mu},\quad
    I_4^\mu = n_{\nu}\sigma^{\nu\mu};
\\
\label{I1-I3-munu}
  &&\!\!\!\!\!\!
    I_{1}^{\mu\nu} = \left(u^\mu\pi^{\nu\lambda}+u^\nu\pi^{\mu\lambda}  
                     \right) Du_{\lambda},\quad 
    I_{2}^{\mu\nu} = \theta \pi^{\mu\nu},\quad 
    I_{3}^{\mu\nu} = \omega^\mu_{\ \lambda}\pi^{\lambda\nu}
                    +\omega^\nu_{\ \lambda}\pi^{\lambda\mu},\qquad
\\
\label{I4-munu}
  &&\!\!\!\!\!\!
    I_{4}^{\mu\nu} = \frac{1}{2}
    \left(\pi^{\mu\lambda}\sigma^{\ \nu}_\lambda
         +\pi^{\nu\lambda}\sigma^{\ \mu}_\lambda\right) 
   -\frac{1}{3}\Delta^{\mu\nu}\pi^{\alpha\beta}\sigma_{\beta\alpha}.
\end{eqnarray}

The conservation laws (\ref{relEqs_Tmut}-\ref{relEqs_n}) together with the dissipative transport equations (\ref{relEqs_Pi}-\ref{relEqs_pi}) constitute the equations of motion of the relativistic hydrodynamic system encoded in \code. Next we will discuss the transport coefficients appearing in these equations.

\subsection{Transport coefficients}
\label{transcoeff}

The EoS and transport coefficients describe the medium properties of the expanding fluid and as such must be determined microscopically. While for the EoS detailed knowledge is available now from lattice QCD (see Sec.~\ref{subsec-eos}), the same is not true for the transport coefficients. We will here use rough estimates for the transport coefficients that have been obtained from kinetic theory, but have to leave their precise determination to future theoretical or phenomenological work. Specifically, \code\ implements the transport coefficients from Ref.~\cite{PhysRevD.85.114047, PhysRevC.90.024912} which starts from the Boltzmann equation in Relaxation Time Approximation (RTA) and employs the 14-moment approximation \cite{PhysRevLett.105.162501, PhysRevD.85.114047} for a one-component gas of Boltzmann particles with non-zero but small mass $m\ll T$. In all expressions we only keep the leading non-zero term in powers of $m/T\ll 1$. This is motivated by the approximate masslessness of the microscopic quark-gluon degrees of freedom that make up the fluid described by \code.

\subsubsection{Shear stress tensor}

For the evolution of the shear stress we take the transport coefficients
\begin{eqnarray}
\frac{\eta}{\tau_\pi} &=& \frac{\ed+\peq}{5}\,, 
\label{etatau}
\\
\frac{\delta_{\pi\pi}}{\tau_\pi} = \frac{4}{3}\,, \qquad
\frac{\tau_{\pi\pi}}{\tau_\pi} &=& \frac{10}{7}\,, \qquad
\frac{\lambda_{\pi\Pi}}{\tau_\pi} = \frac{6}{5}\,.
\label{shearcoeff}
\end{eqnarray}
Following \cite{PhysRevC.81.014902} we express shear viscous effects in terms of the {\it kinematic shear viscosity} 
\begin{equation}
    \bar\eta = \frac{\eta T}{\ed+\peq}\label{etabar}\;.
\end{equation}
For $\mu=0$ the kinematic shear viscosity reduces to the {\it specific shear viscosity} $\eta/\s$, where $\s$ denotes the entropy density, but it differs greatly from $\eta/\s$ at large net baryon densities, and this can lead to significant differences in the hydrodynamic flow patterns \cite{PhysRevC.88.064901}. Parametrizations of $\bar\eta$ as a function of $T$ and $\mu$ are discussed in \cite{NoronhaHostler:2008ju, Denicol:2015nhu}; in \code\ the default setting for $\bar\eta$ is a constant, $\bar\eta=0.2$. Given $\bar\eta$, Eq.~(\ref{etatau}) is used to calculate $\tau_\pi=5\bar\eta/T$, and Eqs.~(\ref{shearcoeff}) to obtain the remaining transport coefficients. 

\subsubsection{Bulk viscous pressure}

To evolve the bulk viscous pressure we take the transport coefficients 
\begin{eqnarray}
\frac{\zeta}{\tau_\Pi}&=&15\left(\frac{1}{3}-c^2_s\right)^{2}(\ed+\peq)\,, 
\label{betaPi}
\\
\frac{\delta_{\Pi\Pi}}{\tau_\Pi}&=&\frac{2}{3}\,,\qquad
\frac{\lambda_{\Pi\pi}}{\tau_\Pi}=\frac{8}{5}\left(\frac{1}{3}-c^2_s\right)\,,
\label{bulkcoeff}
\end{eqnarray}
where $c_s$ is the speed of sound in the medium (see Eq. (\ref{cs2EoS})). Similar to the shear viscosity, we start from a parametrization as a function of $T$ and $\mu$ of the {\it kinematic bulk viscosity}
\begin{equation}
    \bar\zeta = \frac{\zeta T}{\ed+\peq}\label{zetabar}
\end{equation}
and determine from it the bulk relaxation time $\tau_\Pi$ using Eq.~(\ref{betaPi}) and the remaining transport coefficients using Eqs.~(\ref{bulkcoeff}). In the code, we use a parametrization that interpolates between lattice QCD data for the QGP phase and results obtained from the hadron resonance gas model for the hadronic phase, connected quadratically around the pseudocritical temperature $T_c=155$\,MeV \cite{PhysRevC.80.064901}:
\begin{equation}
{\small
  \bar{\zeta} =
  \begin{cases}
  A_0 + A_1 x + A_2 x^2\,, &0.995\,T_c \ge T \ge 1.05\,T_c\,, 
  \\
  \lambda_1 \exp [-(x-1)/\sigma_1] + \lambda_2 \exp [-(x-1)/\sigma_2]+0.001\,,
     &T > 1.05\,T_c\,, 
  \\
  \lambda_3 \exp [(x-1)/\sigma_3] + \lambda_4 \exp [ (x-1)/\sigma_4]+0.03\,,
     &T < 0.995\,T_c\,, 
  \end{cases}
  \label{eq-zetas}
}
\end{equation}
with $x = T/T_c$ and fitted parameters 
\begin{eqnarray*}
  &&A_0=13.45\,,\quad A_1=27.55\,,\quad A_2=-13.77\,, \\
  &&\lambda_1=0.9\,,\quad \lambda_2=0.25\,,\quad\lambda_3=0.9\,,\quad \lambda_4=0.22\,,\\
  &&\sigma_1=0.025\,,\quad \sigma_2=0.13\,, \quad\sigma_3=0.0025\,,\quad \sigma_4=0.022\,.
\end{eqnarray*}

The bulk viscous pressure describes the deviation from the thermal pressure for a non-perfect expanding or contracting fluid. Bulk viscosity has been shown to generate important effects on the slope of the transverse momentum spectra and their azimuthal anisotropy \cite{PhysRevLett.115.132301}. Bulk viscous effects are expected to be strongest near the quark-hadron phase transition \cite{Paech:2006st, Arnold:2006fz, Kharzeev:2007wb, Karsch:2007jc, Meyer:2007dy, Moore:2008ws, Sasaki:2008fg}, especially near the QCD critical point where critically enhanced contributions associated with critical slowing-down play a key dynamical role
\cite{Berdnikov:1999ph, Song:2009rh, PhysRevD.98.036006}.

\subsubsection{Baryon diffusion current}

Compared to the transport coefficients related to the bulk and shear stresses, those controlling baryon diffusion are much less explored. Following Ref.~\cite{Denicol:2018wdp}, we use the coefficients obtained from the Boltzmann equation for an almost massless ($m/T\ll1$) classical gas (for a calculation of the baryon diffusion coefficient $\kappa_n$ for a massive gas of hadrons see Ref.~\cite{Albright:2015fpa}):
\begin{eqnarray}
\label{kappa}
    \kappa_n &=& \tau_n\n\left[\frac{1}{3}\coth\left(\frac{\mu}{T}\right)-\frac{\n T}{\ed+\peq} \right]\,,
\\
\label{diffcoeff}
    \delta_{nn} &=& \tau_n\;,\qquad \lambda_{nn} = \frac{3}{5}\tau_n\;.
\end{eqnarray}
Here $\tau_n$ is the relaxation time of the baryon diffusion current in Eq.~(\ref{eq-relax-n}) and parametrized as 
\begin{equation}
\label{taun}
    \tau_n = \frac{C_B}{T},
\end{equation}
with a free parameter $C_B$.\footnote{%
        Note that this procedure is different from the bulk and shear viscosity where we parametrized the first-order transport coefficients and computed the relaxation times from them, whereas here we parametrize the relaxation time and use it to compute the diffusion coefficient.}
The expression (\ref{kappa}) for the diffusion coefficient $\kappa_n$ was derived in first-order Chapman-Enskog approximation \cite{Denicol:2018wdp} whereas the second-order transport coefficients (\ref{diffcoeff}) were obtained in the 14-moment approximation \cite{PhysRevC.90.024912, PhysRevLett.105.162501, PhysRevD.85.114047}. In the limit of small net baryon density ($\mu\to0$) the diffusion coefficient $\kappa_n$ reduces to
\begin{equation}
\label{kappa_small_mu}
    \frac{\kappa_n}{\tau_n} = \frac{\n T}{3\mu}\;.
\end{equation}

In RTA, where the collision term in the Boltzmann equation is parametrized with a single relaxation time $\tauR$, the relaxation times for $\Pi$, $\pi^{\mu\nu}$ and $\V^\mu$ are all the same, i.e. $\tau_\Pi=\tau_\pi = \tau_n = \tauR$. Here we allow them to be different. If one does, however, impose the constraint $\tau_n=\tau_\pi=\tauR$, the baryon diffusion and shear viscosity coefficients can be related as follows \cite{Jaiswal:2015mxa}:
\begin{equation}
\label{kappaeta}
    \frac{\kappa_n T}{\eta} = C\left(\frac{\mu}{T}\right)\,
    \left(\frac{\pi T}{\mu}\right)^2
    \left(\frac{\n T}{\ed+\peq}\right)^2\,.
\end{equation}
In contrast to Eqs.~(\ref{kappa},\ref{kappa_small_mu}), this expression takes into account quantum statistics. The function $C(\mu/T)$ exhibits a weak dependence on $\mu/T$, interpolating between 5/3 at large $\mu/T$ and a somewhat smaller value at small $\mu/T$ whose precise magnitude depends on the number of massless degrees of freedom in the gas \cite{Jaiswal:2015mxa}. Note that, since for small $\mu/T$ the net baryon density $\n\propto\mu$, both (\ref{kappa_small_mu}) and (\ref{kappaeta}) yield nonzero baryon diffusion coefficients at zero net baryon density. At large $\mu/T$, the authors of Ref.~\cite{Jaiswal:2015mxa} have shown that the ratio $\kappa_n T/\eta$ approaches zero, i.e., at large net baryon densities and low temperatures baryon diffusion effects can generally be neglected in comparison with shear viscous stresses. 

In addition to kinetic theory, gauge/gravity duality has also been widely used to determine the transport properties of the QGP (see, for example, \cite{PhysRevD.77.066014, Rougemont:2015ona, Son:2006em}). With this method it is possible to study the transport properties of strongly coupled gauge theories for which no kinetic theory description exists, including the critical dynamics near a critical point. The latter is expected to provide critical signatures for the experimental identification of the QCD critical point \cite{Berdnikov:1999ph, PhysRevD.98.036006}. Studying and comparing the baryon number evolution with baryon diffusion coefficients corresponding to a weakly \cite{Denicol:2018wdp} or a strongly coupled QGP \cite{Rougemont:2015ona} can be interesting \cite{Du:2018mpf, PhysRevC.98.064908}.

\subsection{Equation of State (EoS)}\label{subsec-eos}

Another important medium property that crucially affects the dynamical evolution of the fluid is its equation of state 
\begin{equation}
    \peq = \peq(\ed, \n) = \peq(T,\mu)\,.
\label{eq-eos}
\end{equation}
In practice, for the calculation of the transport coefficients and chemical forces we also need the equivalent relations
$T(\ed,\n)$ and $\mu(\ed,\n)$. We use the term EoS generically for any one of these relations.

\subsubsection{Construction of the EoS}
\label{subsec-coneos}

Since the matter produced in nuclear collisions passes through very different physical regimes that differ by orders of magnitude in energy density and must be described with different effective degrees of freedom, we need an EoS that describes the medium properties over a wide range of temperature and length scales \cite{PHILIPSEN201355}. On most scales the degrees of freedom of the evolving system are strongly coupled, rendering  perturbative investigations from first principles unreliable. Over the last decades, lattice QCD (LQCD) has been established as the most precise non-perturbative framework to calculate the EoS of strongly interacting matter at zero baryon chemical potential (see, e.g., \cite{Borsanyi:2010cj}). The method works well at temperatures above $\sim 100$\,MeV; at lower temperatures the lattice signals become weaker and more noisy, necessitating the matching of LQCD data to an analytical hadron resonance gas model.   

Lattice QCD obtains the EoS by calculating the trace of the energy-momentum tensor $T^{\mu\nu}$, $\ed - 3\peq$ (usually referred to as the ``trace anomaly'' or ``interaction measure''), describing deviations from the conformal EoS. Defining the rescaled dimensionless trace anomaly
\begin{equation}
    I(T, \mu) \equiv \frac{\ed(T, \mu) - 3\peq(T, \mu)}{T^4}\;,
\end{equation}
the thermal pressure at zero chemical potential can be written as
\begin{equation}
    \frac{\peq^\mathrm{LAT}(T, 0)}{T^4} = \int^T_0 dT'\frac{I^\mathrm{LAT}(T', 0)}{T'}\;.
\end{equation}

Unfortunately, this method cannot be directly extended to non-zero chemical potential where the evaluation of the QCD path integral for the interaction measure $I(T,\mu)$ suffers from a ``sign problem'' \cite{PHILIPSEN201355}, precluding its direct computation with standard Monte-Carlo methods. This problem can be partially circumvented by using standard LQCD techniques to also compute the $\mu$-derivatives of the pressure $\peq(T,\mu)$ at $\mu=0$ and construct $\peq(T,\mu)$ at non-zero $\mu$ from its Taylor series around $\mu=0$:
\begin{equation}
    \frac{\peq^\mathrm{LAT}(T, \mu)}{T^4} = \frac{\peq^\mathrm{LAT}(T, 0)}{T^4} + \sum_{n=1}^{n_\mathrm{max}} c_{2n}(T)\left(\frac{\mu}{T}\right)^{2n}\,.
\label{Taylor}
\end{equation}
The expansion coefficients 
\begin{equation}
    c_n(T) = \left.\frac{1}{n!}\frac{\partial^n (\peq/T^4)}{\partial(\mu/T)^n}\right|_{\mu=0}
\end{equation}
are known as the ``baryon number susceptibilities'' \cite{Borsanyi2012, Parotto:2018pwx, PhysRevD.95.054504}.
The computational effort of computing them increases rapidly with their order $n$; at this time, the Taylor expansion (\ref{Taylor}) includes terms up to order $n_\mathrm{max}=3$ and, near $T=T_c$, converges well up to about baryon chemical potentials $\mu/T \lesssim 2$ \cite{Borsanyi2012, Parotto:2018pwx, PhysRevD.95.054504}. 

At low temperatures $T\ll T_c\simeq 155$\,MeV, LQCD is increasingly affected by lattice artifacts and the system is more properly described in terms of hadronic degrees of freedom
as a ``hadron resonance gas'' (HRG). In the HRG model, the interactions among different hadronic species are accounted for by including all experimentally identified scattering resonances as additional, non-interacting particle species. In the HRG model the interaction measure $(\ed{-}3\peq)/T^4$ is given as  \cite{PhysRevD.90.094503, HUOVINEN201026}
\begin{equation}
    I^\mathrm{HRG}(T, 0)=\sum_{m_i\leq m_\mathrm{max}}\frac{g_i}{2\pi^2}\sum^\infty_{k=1}\frac{(-\eta_i)^{k+1}}{k}\left(\frac{m_i}{T}\right)^3K_1\left(\frac{km_i}{T}\right),
\end{equation}
where particle species with spin-isospin degeneracy $g_i$ and mass $m_i$ smaller than some cut-off $m_\mathrm{max}$ can be included, and $\eta_i = -1~(+1)$ for bosons (fermions) describes the effects of quantum statistics. 

In principle, the EoS used for the hydrodynamic evolution should include the same set of hadronic resonances as the hadronic afterburner employed to describe the kinetic final freeze-out stage because otherwise a mismatch of the energy and baryon densities occurs on the conversion surface where we change between these two different dynamical descriptions. In practice these discontinuities tend to be small, and such care is not always taken. In our applications of BEShydro we use different matched equations of state for different hadronic afterburners (e.g., for {\sc UrQMD} \cite{Bass:1998ca, Bleicher:1999xi} and {\sc SMASH} \cite{Weil:2016zrk}); however, the module for matching the lattice QCD data to a HRG with adjustable hadronic mass spectrum is not part of this code distribution. 

For the matching procedure between the LQCD and HRG equations of state different methods have been used. For example, in Refs. \cite{Parotto:2018pwx, Denicol:2018wdp} the pressure is interpolated as follows:
\begin{eqnarray}
    \frac{\peq(T, \mu)}{T^4} &=& \frac{1}{2}
    \left[1-\tanh\left(\frac{T{-}T'(\mu)}{\Delta T'}\right)\right]
    \frac{\peq^\mathrm{HRG}(T, \mu)}{T^4}
\\
    &+& \frac{1}{2}
    \left[1+\tanh\left(\frac{T{-}T'(\mu)}{\Delta T'}\right)\right]
    \frac{\peq^\mathrm{LAT}(T, \mu)}{T^4}\,.
\end{eqnarray}
Here $\peq^\mathrm{HRG}$ and $\peq^\mathrm{LAT}$ are the equilibrium pressures for the hadron resonance gas and from lattice QCD, respectively, $T'(\mu)$ is the ``switching temperature'' and $\Delta T'$ controls the width of the ``overlap region''. The authors of Ref.~\cite{PhysRevC.93.044913}, on the other hand, interpolate the interaction measure at $\mu=0$, $I(T,\mu=0)$ smoothly between $T_1=T_c(\mu{=}0)=155$\,MeV and $T_2=180$\,MeV, using a polynomial interpolation function. 

Once the pressure $\peq(T, \mu)$ is given, other thermodynamic quantities can be calculated from thermodynamic identities:
\begin{eqnarray}
    \frac{\s(T, \mu)}{T^3} &=& \frac{1}{T^3}\left[\frac{\partial \peq(T, \mu)}{\partial T}\right]_{\mu}\;,
\label{sEoS}\\
    \frac{\n(T, \mu)}{T^3} &=& \frac{1}{T^3}\left[\frac{\partial \peq(T, \mu)}{\partial \mu}\right]_T\;,
\label{nEoS}\\
    \frac{\ed(T, \mu)}{T^4} &=& \frac{\s(T, \mu)}{T^3}-\frac{\peq(T, \mu)}{T^4}+\frac{\mu}{T}\frac{\n(T, \mu)}{T^3}\;,
\label{eEoS}\\
    c^2_s(T, \mu) &=& \left[\frac{\partial \peq(\ed, \n)}{\partial \ed}\right]_{\n} + \frac{\n}{\ed+\peq}\left[\frac{\partial \peq(\ed, \n)}{\partial \n}\right]_{\ed}\;.
\label{cs2EoS}
\end{eqnarray}
The last equation requires expressing $\ed$ and $\n$ through $T$ and $\mu$ after taking the derivatives. In practice, the functions $\ed(T,\mu)$ and $\n(T,\mu)$ are numerically inverted, and the quantities $T, \mu, \s, \peq$, and $c^2_s$, as well as the two derivatives on the r.h.s. of Eq.~(\ref{cs2EoS}), are stored in a table on a $(\ed,\n)$ grid which is interpolated by the hydrodynamic code as needed. 

\subsubsection{Equations of state implemented in \code}

In \code\ four different equations of state, EOS1 to EOS4, are implemented, for different purposes: at zero chemical potential, we include a massless (conformal) EoS ($\ed=3\peq$, EOS1) as well as an interpolated LQCD-HRG EoS from the Wuppertal-Budapest collaboration \cite{Borsanyi:2010cj} (EOS2); at non-zero chemical potential, an appropriately generalized conformal EoS (EOS3) and an interpolated LQCD-HRG EoS from Ref.~\cite{Denicol:2018wdp} (EOS4) are used.

EOS1 assumes an ideal gas of massless quarks and gluons:
\begin{equation}
        \ed = 3\peq = 3\left[2(N_c^2-1)+\frac{7}{2}N_cN_f\right]
                      \frac{\pi^2}{90}T^4\;,
\label{EOS1}
\end{equation}
where $N_c = 3$ and $N_f = 2.5$ are the numbers of colors and (approximately) massless quark flavors, respectively.\footnote{%
    Strange quarks, whose mass is of the same order of magnitude as the quark-hadron transition temperature, are (somewhat roughly) counted as 1/2 massless quark flavor.} 
In EOS1, $c_s^2=1/3$ and $\mu = 0$. While the conformal EoS does not properly describe the properties of the matter produced in nuclear collisions, it is, owing to its simplicity, very useful for code testing. Technically, EOS1 can be used in \code\ even when the baryon density and baryon diffusion currents are being evolved; in that case, these currents do not couple to the rest of the hydrodynamic system and evolve purely as background fields.

\begin{figure*}[!htbp]
\centering
    \begin{subfigure}
        \centering
        \includegraphics[width=0.45\textwidth]{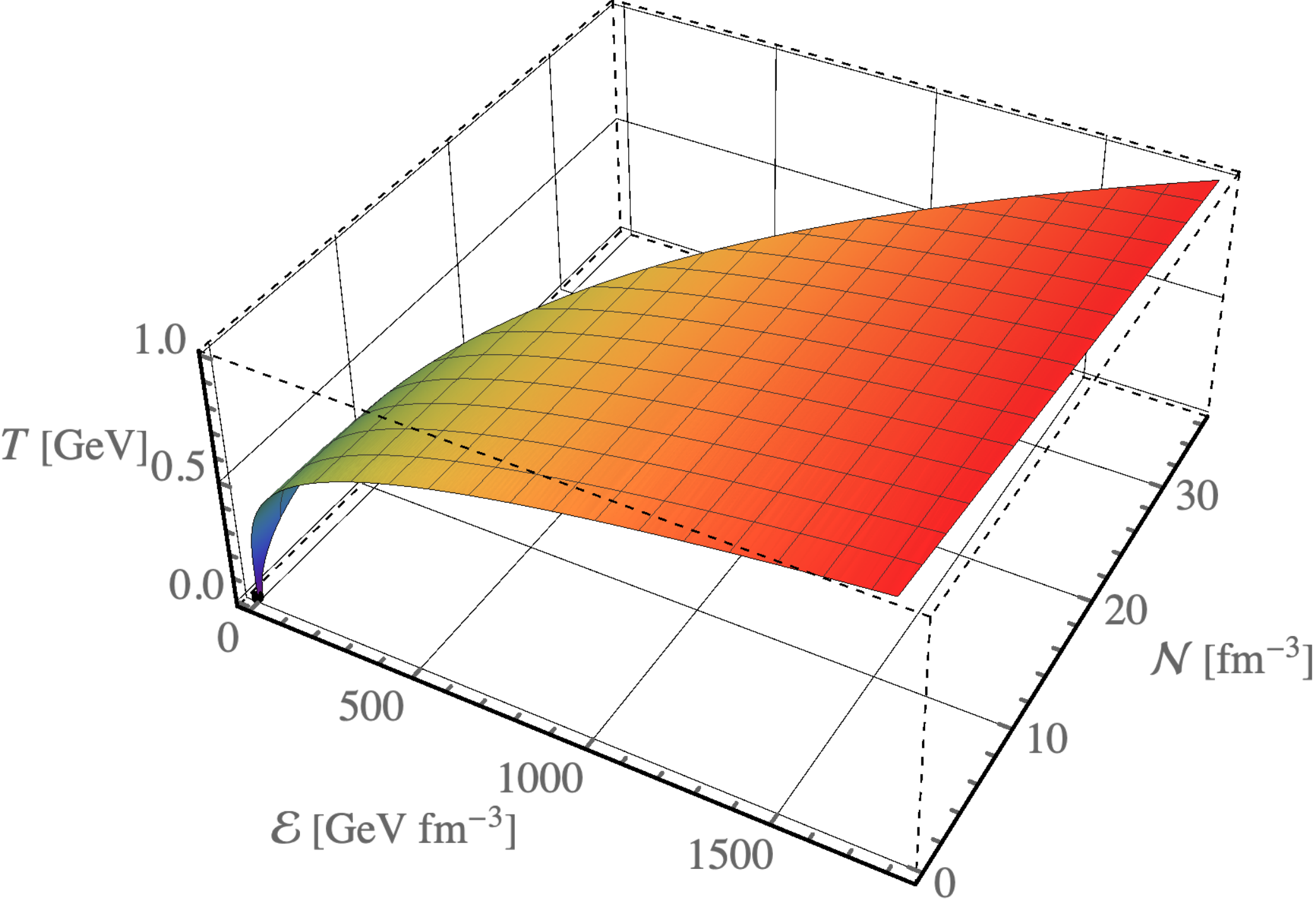}
    \end{subfigure}
    \begin{subfigure}
        \centering
        \includegraphics[width=0.45\textwidth]{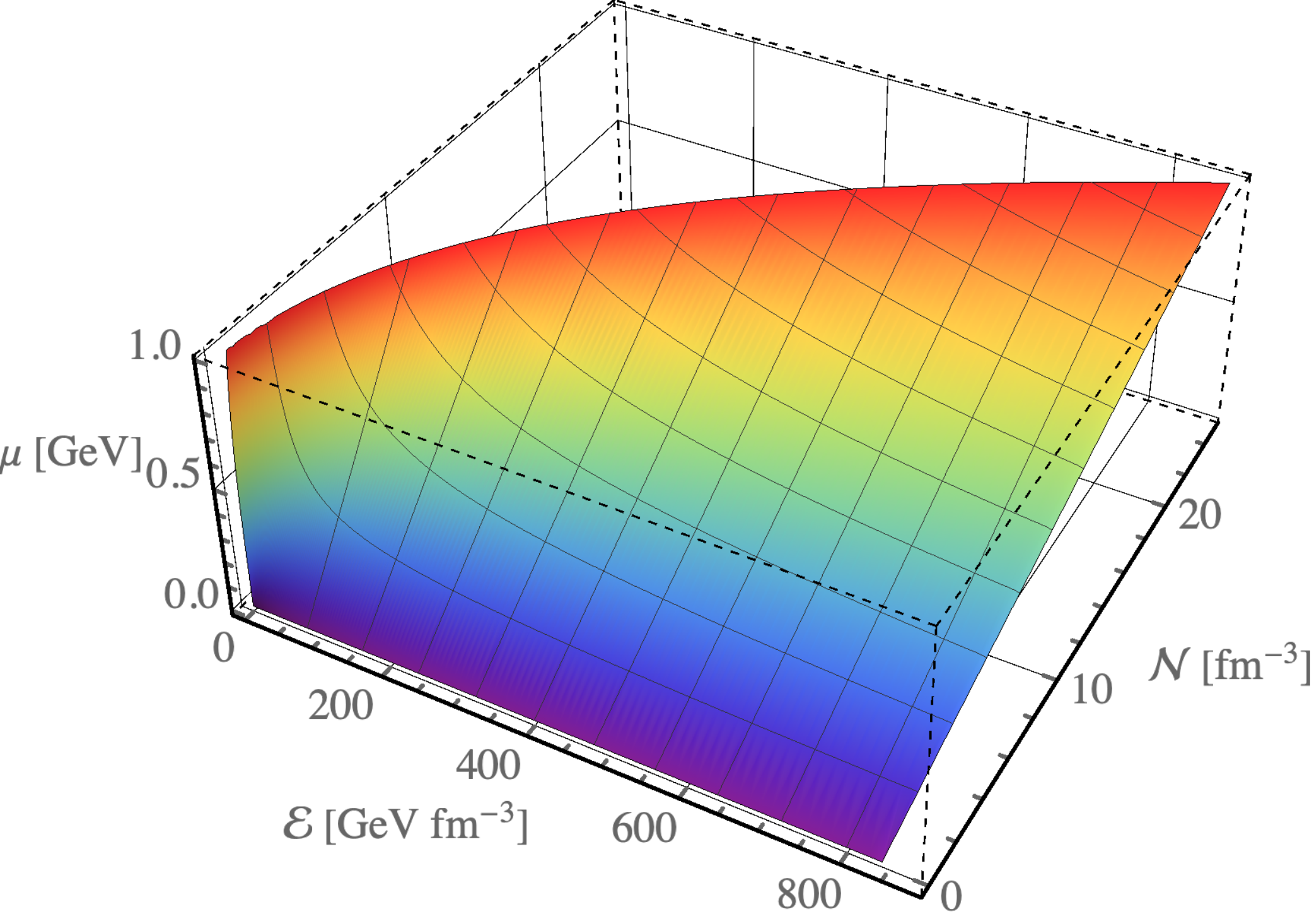}
    \end{subfigure}
    \caption{(Color online) Conformal EOS3 with $N_f=2.5$ at finite temperature and chemical potential. {\sl Left:} $T(\ed, \n)$. {\sl Right:} $\mu(\ed, \n)$.
    }
    \label{F1}
\end{figure*}

EOS2 from the Wuppertal-Budapest collaboration \cite{Borsanyi:2010cj} can be used for realistic simulations at vanishing $\mu$, i.e. for heavy-ion collisions at ultra-relativistic collision energies with $\sqrt{s_\mathrm{NN}} \gg 100$\,GeV, especially near mid-rapidity. More details about EOS2 can be found in Refs.~\cite{Borsanyi:2010cj, Bazow:2016yra}.

EOS3 is the generalization of EOS1 to non-zero $\mu$. Starting from the ideal massless parton gas expression \cite{Mueller2018}
\begin{equation}
    \frac{\peq}{T^4} = \frac{\pi^2}{90}\left[2(N_c^2-1)+\sum_f 4N_c\left(\frac{7}{8}+\frac{15}{4}\left(\frac{\mu_f}{\pi T}\right)^2+\frac{15}{8}\left(\frac{\mu_f}{\pi T}\right)^4\right)\right],
\label{EOS3}
\end{equation}
%
%
\begin{figure*}[!b]
\centering
    \begin{subfigure}
        \centering
        \includegraphics[width=0.45\textwidth]{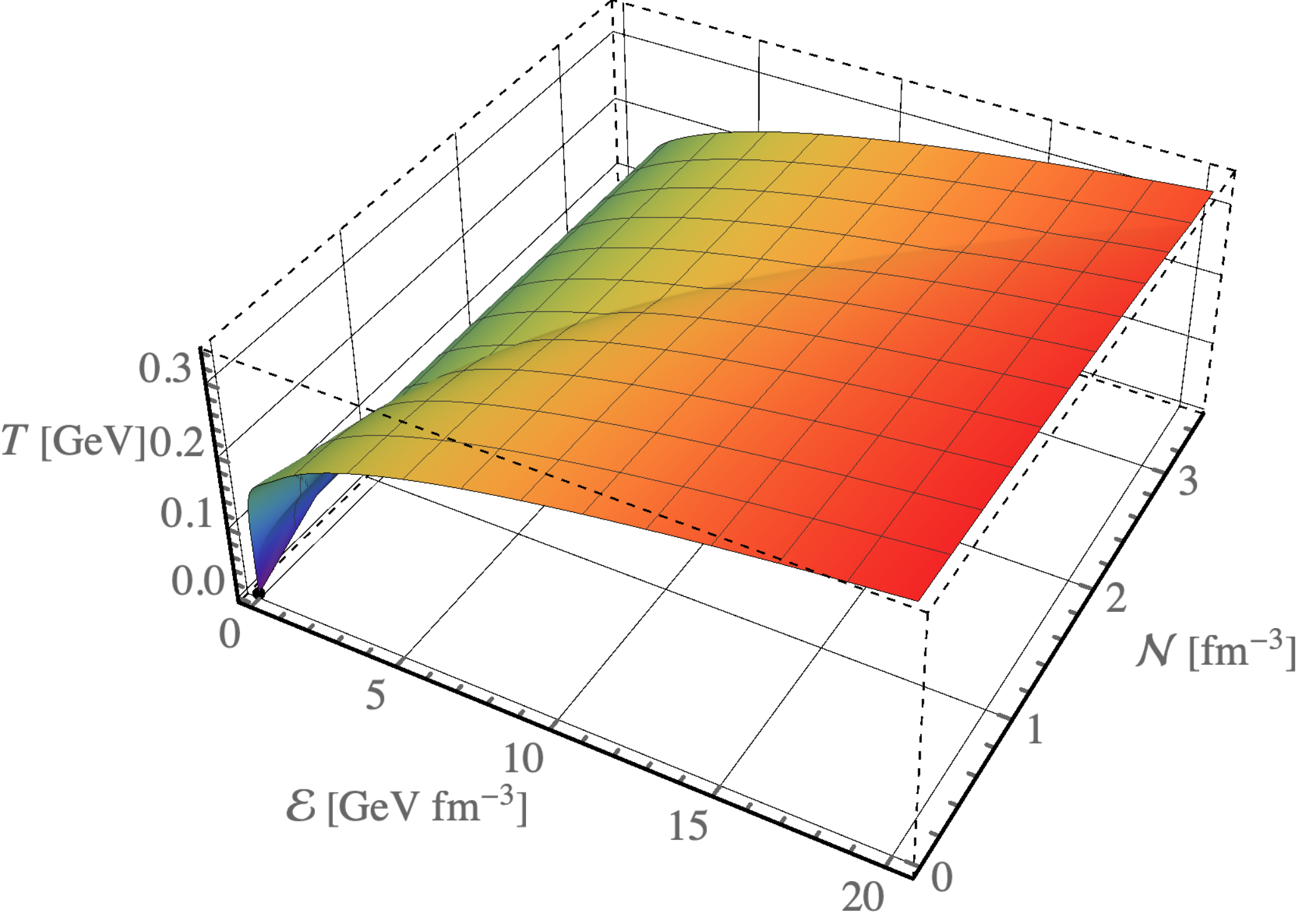}
    \end{subfigure}
    \begin{subfigure}
        \centering
        \includegraphics[width=0.45\textwidth]{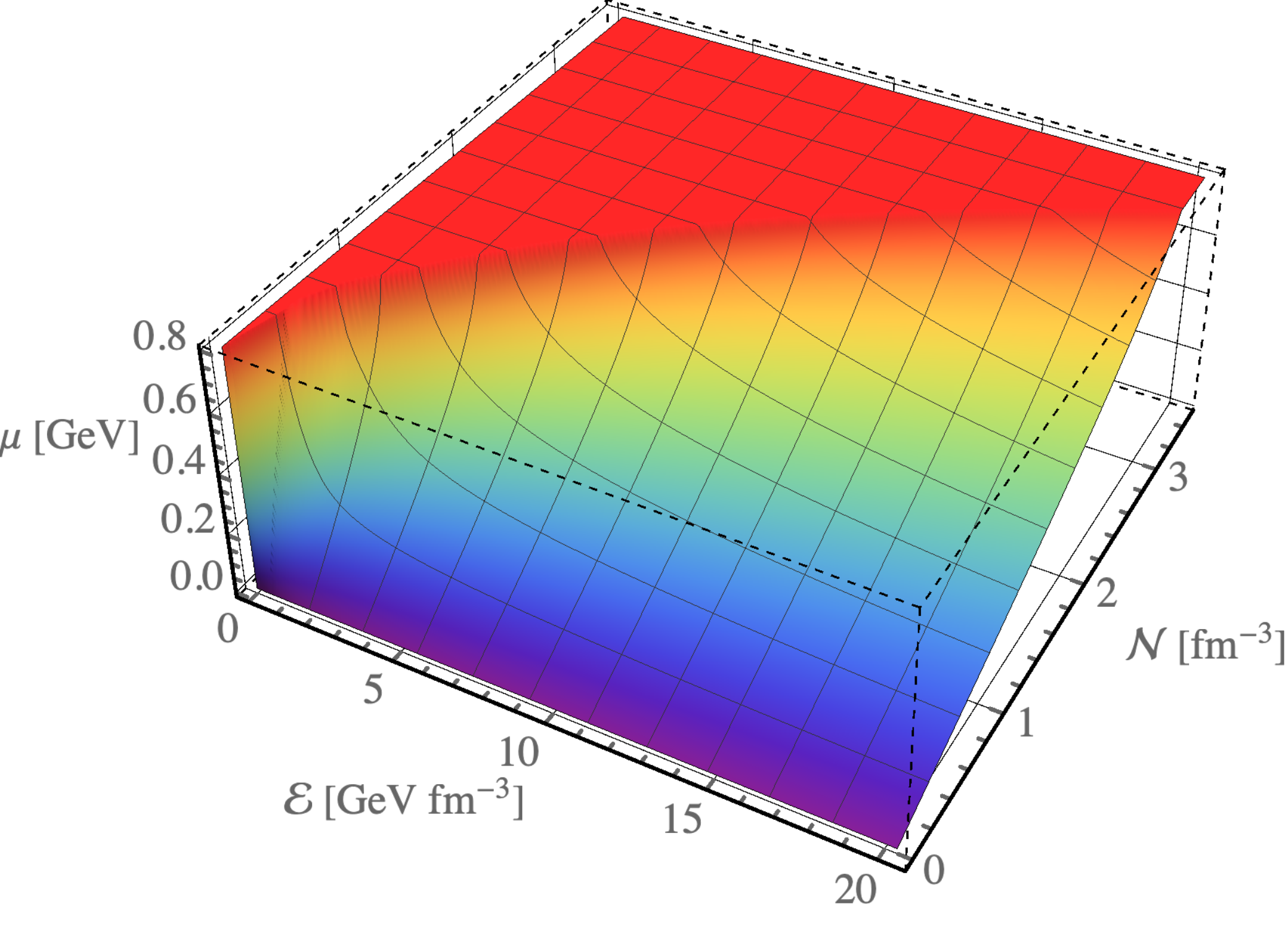}
    \end{subfigure}
    \caption{(Color online) EOS4 from Ref.~\cite{Denicol:2018wdp}, restricted to $\mu<800$\,MeV to account for the limited range of validity of the Taylor series extrapolation to non-zero $\mu$ ({\sl left:} $T(\ed, \n)$, {\sl right:} $\mu(\ed, \n)$). In the flat region of the left plot, as $\n$ is increased beyond its edge $\n_\mathrm{edge}(\ed)$, $T(\ed,\n)$, $\mu(\ed,\n)$ and $\peq(\ed,\n)$ are set by hand to remain constant (i.e. $T(\ed,\n>\n_\mathrm{edge}(\ed)) = T(\ed,\n_\mathrm{edge}(\ed))$, etc.). (Note that in the flat region the entropy density $\s$ can go negative if naively calculated from $\s=(\ed+\peq-\mu\n)/T$. Correspondingly, this EOS should not be used for collision systems for which the code makes regular excursions into this region.) The same prescription is then also used in the right plot.
    \label{F2}
    }
\end{figure*}%
%
where $N_c=3$ and the sum goes over massless quark flavors, we simplify it by setting $\mu_f = \mu/3$ for all flavors (which is appropriate if only baryon number is considered as a conserved charge): 
\begin{eqnarray}
  \frac{\peq(T,\mu)}{T^4}&=&
  p_0+N_f\left[\frac{1}{18}\left(\frac{\mu}{T}\right)^2
     +\frac{1}{324\pi^2}\left(\frac{\mu}{T}\right)^4\right]
     =\frac{\ed(T,\mu)}{3T^4}\;,
\label{EOS3P}
\\
  \frac{\n(T,\mu)}{T^3}&=&
   N_f\left[\frac{1}{9}\left(\frac{\mu}{T}\right)
           +\frac{1}{81\pi^2}\left(\frac{\mu}{T}\right)^3\right]\,,
\label{EOS3N}
\end{eqnarray}
with $p_0=(16+10.5N_f)\pi^2/90$. We again count strange quarks with a factor 1/2, i.e. we set $N_f=2.5$ so that for $\mu=0$ (\ref{EOS3P}) reduces to (\ref{EOS1}). Inverting these functions numerically one obtains the EoS tables used in the hydrodynamic code (see Fig.~\ref{F1}).

EOS4 from Ref.~\cite{Denicol:2018wdp}, extended to finite baryon chemical potential by combining a lattice EoS at high temperature and a HRG EoS at low temperature with a Taylor expansion in $\mu/T$ using techniques discussed in Sec.~\ref{subsec-coneos}, allows to study the evolution of systems with non-zero net baryon density. It is plotted in Fig.~\ref{F2}. In the code, tabulated values for $\peq,\, T$ and $\mu/T$ as functions of $(\ed, \n)$ are included. If the code requires the EoS at $(\ed, \n)$, $\peq, T$ and $\mu/T$ are calculated from nearest neighbors in the table using 2D bilinear interpolation. We note that EOS4 does not include a critical point or first-order phase transition at large $\mu$. A lattice QCD based EoS that includes these features, with adjustable location of the critical point and strength of the first-order transition beyond that point, was constructed by the BEST Collaboration \cite{Parotto:2018pwx} and could be imported into \code\ for future dynamical simulations aiming at helping to locate the QCD critical point.

\section{Numerical scheme}
\label{numericalscheme}

We now describe the numerical scheme used in \code\ to solve the coupled set of evolution equations (the conservation laws (\ref{relEqs_Tmut}-\ref{relEqs_n}) together with the dissipative relaxation equations (\ref{relEqs_Pi}-\ref{relEqs_pi}) and the EoS (\ref{eq-eos})) discussed in Section~\ref{phys-framework}. Initial values for all components of the baryon charge current and energy-momentum tensor are set on a surface of constant longitudinal proper time $\tau$.\footnote{%
        A dynamical initialization routine with sources for the divergences of the baryon current and energy-momentum tensor that describe the gradual ``hydrodynamization'' of the matter produced in the collision \cite{Shen:2017bsr, Akamatsu:2018olk, Du:2018mpf}
        will be discussed elsewhere.}
We focus our attention on aspects of the algorithm related to the evolution and influence of the baryon density and diffusion currents, referring interested readers to Refs. \cite{PhysRevC.82.014903, Molnar:2009tx, Bazow:2016yra} for additional technical details.

\subsection{The Kurganov-Tadmor algorithm}
\label{KT}

Using the definition of $d \equiv u^\mu \partial_\mu$, the equations of motion in Eqs. (\ref{relEqs_Tmut}-\ref{relEqs_n}), (\ref{relEqs_Pi}-\ref{relEqs_pi}) can all be written \cite{Molnar:2009tx, Bazow:2016yra} in the same first-order flux-conserving form
\begin{equation}
    \partial_\tau q + \partial_x (v^x q) + \partial_y (v^y q) + \partial_\eta (v^\eta q) = S_q\;,
\label{numericalEquations}
\end{equation}
where $v^i \equiv u^i/u^\tau$ ($i = x, y, \eta_s$) is the 3-velocity of the fluid, the conserved quantity $q$ can be any component (or linear combination of components) of $T^{\mu\nu}$ and $N^{\mu}$, and $S_q$ is the corresponding source term. The
explicit equations (\ref{numericalEquations}) implemented in \code\ are given in Appendix~\ref{appa}. This form allows all quantities to be evolved with the same numerical transport algorithm.

\code\ is designed for flexibility so that different physical limits can be easily studied. Only the propagation of $T^{\tau\tau}$, $T^{\tau x}$, $T^{\tau y}$, and $T^{\tau \eta}$ is always turned on. The evolution of all other variables (dissipation and/or baryon related) can be conveniently turned on and off independently; only the propagation of $\V^\mu$ requires the evolution of $N^{\tau}$ to be turned on. 

To solve equation (\ref{numericalEquations}) we use the Kurganov-Tadmor (KT) algorithm \cite{KURGANOV2000241}, with a second-order explicit Runge-Kutta (RK) ordinary differential equation solver \cite{leveque_2002} for the time integration step. This scheme is widely used in relativistic hydrodynamic simulations (see, e.g., \cite{PhysRevC.82.014903, Pang:2018zzo, Bazow:2016yra}). 

\subsection{Numerical derivatives}

For the source terms $S_q$ we must evaluate spatial and temporal derivatives of the thermodynamic variables and dissipative flows. For the time derivatives the code uses first-order forward differences:
\begin{equation}
    \partial_{\tau}A^{n}_{i,j,k}=\frac{A^{n}_{i,j,k}-A^{n-1}_{i,j,k}}{\Delta\tau}\,. 
\label{eq-time-de}
\end{equation}
Here $A$ is the quantity to be differentiated, $i,j,k$ are  integer labels for the $x, y$, and $\eta_s$ coordinates of the grid point, and $\Delta \tau$ is the temporal grid size (numerical resolution in the $\tau$ coordinate). $n$ and $n-1$ are temporal indices denoting the present and preceding time step. To initialize the temporal evolution code at the first time step $n=1$ we set $A^{0}_{i,j,k} = A^{1}_{i,j,k}$. This is especially important when the initial flow velocity $u^\mu$ is non-zero, for example in the case of the Gubser flow test described in Sec.~\ref{gubsertest}.

The code provides two methods for calculating spatial derivatives. The first uses second-order central differences, i.e. the derivative of any quantity $A$, say, with respect to $x$ is calculated as
\begin{equation}
    \partial_{x}A^n_{i,j,k} = \frac{A^n_{i+1,j,k}-A^n_{i-1,j,k}}{2\Delta x},\label{eq-cendiff}
\end{equation}
where $\Delta x$ is the numerical resolution (grid size) in $x$ direction. The boundary conditions are taken care of by introducing ghost cells on the boundary as described in \cite{Bazow:2016yra}.

The second method calculates the spatial derivative from a combination of second-order central and first-order backward and forward derivatives, using the generalized minmod flux limiter:
\begin{equation}
    \partial_{x}A^n_{i,j,k} = \mathrm{\texttt{minmod}}\left(
\theta_\mathrm{f}\frac{A^n_{i,j,k} - A^n_{i-1,j,k}}{\Delta x},\,
\frac{A^n_{i+1,j,k} - A^n_{i-1,j,k}}{2\Delta x},\,
\theta_\mathrm{f}\frac{A^n_{i+1,j,k} - A^n_{i,j,k}}{\Delta x}
\right)\;,
\label{eq-approx-derivative}
\end{equation}
where the multivariate minmod function is defined as
\begin{align}
\mathrm{\texttt{minmod}}(x,y,z)\equiv \mathrm{\texttt{minmod}}(x,\mathrm{\texttt{minmod}}(y,z))\,,
\end{align}
with $\mathrm{\texttt{minmod}}(x,y)\equiv[\mathrm{\texttt{sgn}}(x)+\mathrm{\texttt{sgn}}(y)]\cdot
\mathrm{\texttt{min}}(|x|,|y|)/2$ and $\mathrm{\texttt{sgn}}(x)\equiv |x|/x $. In other words, $\mathrm{\texttt{minmod}}(x,y,z)$ always gives the value which is the closest to 0 among $(x,y,z)$.  The parameter $\theta_\mathrm{f}\in[1,2]$; $\theta_\mathrm{f}=1$ ($\theta_\mathrm{f}=2$) corresponds to the most (least) dissipative limiter. In \code\ Eq.~(\ref{eq-approx-derivative}) is used only for the derivatives of $u^\mu$ and $\peq$, and only when selected by the user as an option.

\subsection{Root finding with baryon current}
\label{sec-root}

The code evolves the components of the energy momentum tensor in the global computational frame, but the EoS (which is needed to close the set of evolution equations) and the computation of the source terms on the r.h.s. of Eq.~(\ref{numericalEquations}) require knowledge of fluid velocity $u^\mu$ and the energy and baryon density in the local rest frame of the fluid. Computing the latter from the former is known as the ``root finding'' problem. This must be done as efficiently as possible since this problem must be solved at every point of the computational space-time grid.

At finite baryon density, with nonzero baryon diffusion current, the root finding algorithm becomes more complex than described in Ref. \cite{Bazow:2016yra}. We here describe the most general form of the root finding problem: assuming that $T^{\tau\mu}$, $N^{\tau}$, $\pi^{\tau\mu}$, $\Pi$, and $n^\tau$ are all known from the latest temporal update step, we want to compute $\ed$, $\n$,  and $u^{\mu}$. As will be demonstrated in Sec.~\ref{numericaltests}, the following algorithm \cite{Karpenko:2013wva, Shen:2014vra, Pang:2018zzo} works for both ideal and dissipative fluids, i.e. for both vanishing and non-vanishing dissipative flows. We start by introducing the ``ideal fluid contributions'' $M^\mu$ and $J^\tau$ to the energy-momentum current $T^{\tau\mu}$ and baryon density $N^\tau$ in the computational frame:
\begin{align}
  M^{\tau} &= T^{\tau\tau}-\pi^{\tau\tau} = ( \ed+\p) ( u^{\tau} )^{2} -\p\;,\label{eq-mtau}\\
  M^{i} &= T^{\tau i}-\pi^{\tau i} = ( \ed+\p) u^{\tau} u^{i} \quad (i = x, y, \eta_s)\;,\label{eq-mi}\\
  J^\tau &= N^{\tau}-n^\tau =\n u^{\tau}\,.
\label{eq-jt}
\end{align}
Note that for a viscous fluid $\p = \peq + \Pi$ includes implicitly the bulk viscous pressure. We use the following only when baryon evolution is turned on; otherwise, we use the simpler algorithm described in \cite{Bazow:2016yra} where $\ed$ is found first, using a 1-dimensional zero search. Here we first find the magnitude of the flow velocity, $v$, by solving iteratively \cite{Karpenko:2013wva, Shen:2014vra, Pang:2018zzo}
\begin{equation}
    v \equiv \frac{M}{M^\tau +\p} = 
    \frac{M}{M^\tau +\peq\bigl(\ed(v),\n(v)\bigr) + \Pi}\,,
\label{root-v-expression}
\end{equation}
where $M \equiv \sqrt{(M^x)^2+(M^y)^2+\tau^2(M^\eta)^2}$ and $\ed(v)$, $\n(v)$ are obtained from the known quantities $M^\tau$, $M$, and $J^\tau$ as
\begin{align}
  \ed(v) &= M^\tau - vM\,,
\label{root-e}\\
  \n(v) &= J^\tau\sqrt{1-v^2}\,.
\label{root-n}
\end{align}
Once the flow magnitude $v$ is known, we also know the flow 4-velocity:
\begin{align}
  u^\tau &=\frac{1}{\sqrt{1-v^2}}\,,
\label{eq-utau}\\
  u^i &= u^\tau\frac{M^i}{M^\tau + \p}\,,
\end{align}
for $i = x,y,\eta_s$. Note that the algorithm makes active use of all the numerically known components listed above. 

Rewriting Eq.~(\ref{root-v-expression}) in the form
\begin{equation}
    f(v) \equiv v-\frac{M}{M^\tau+\peq(\ed(v),\n(v))+\Pi}
    =0\,,
\label{root-v-function}
\end{equation}
we solve it by the standard Newton-Raphson method, by repeatedly updating the velocity with
\begin{equation}
    v_{i+1} = v_i - \frac{f\bigl(v_i\bigr)}{f'\bigl(v_i\bigr)}\,,
\label{eq-root-vnewton}
\end{equation}
where
\begin{equation}
    f'(v) \equiv \frac{\partial f(v)}{\partial v} = 1+\frac{M}{(M^\tau+\p)^2}\frac{d \peq}{d v}\,,
\label{dfdv}
\end{equation}
until a sufficiently accurate value is reached. The last term in (\ref{dfdv}) is evaluated numerically using 
\begin{equation}
    \frac{d \peq}{d v}=-\left[M\frac{\partial \peq}{\partial \ed}+J^\tau \frac{v}{\sqrt{1-v^2}}\frac{\partial \peq}{\partial \n}\right]\;, \label{eq-pvderivative}
\end{equation}
where (for equations of state like EOS4) the derivatives $\partial \peq/\partial \ed$ and $\partial \peq/\partial \n$ must be interpolated from the values stored in the EoS table to the pair $(\ed,\n)$ tried in each step of the iteration. 

When $v$ gets close to the speed of light, the Newton-Raphson iteration must be modified to avoid excursions into the causally forbidden region $v>1$. This can lead to numerical instabilities and/or poor convergence. We therefore follow the recipe proposed in \cite{Shen:2014vra} and use Eq.~(\ref{root-v-function}) to solve for $v$ only if in the previous time step, at the spatial grid point in question, $v \leq 0.563624$ or, equivalently (see (\ref{eq-utau})), $u^\tau \leq 1.21061$. Otherwise we instead solve for $u^\tau$ (which has no upper limit), by employing the Newton-Raphson algorithm to find the zero of
\begin{equation}
    f(u^\tau) \equiv u^\tau - \sqrt{\frac{M^\tau+\peq\bigl(\ed(u^\tau),\n(u^\tau)\bigr)+\Pi}
         {\ed+\peq\bigl(\ed(u^\tau),\n(u^\tau)\bigr)+\Pi}}\,.
\label{root-u-function}
\end{equation}
In this case we need to evaluate in each iteration 
\begin{equation}
    f'(u^\tau) = 1 - \frac{1}{2}\left[\frac{\ed-M^\tau}{(\ed+\p)^{3/2}(M^\tau+\p)^{1/2}}\right]\frac{d\peq}{du^\tau}\;,
\end{equation}
where
\begin{equation}
    \frac{d\peq}{du^\tau} = -\frac{1}{(u^\tau)^2}\left[\frac{M}{vu^\tau}\frac{\partial \peq}{\partial \ed} + J^\tau\frac{\partial \peq}{\partial \n}\right]\,,
\label{eq-puderivative}
\end{equation}
with $v = \sqrt{1-1/(u^\tau)^2}$. In Sec.~\ref{gubsertest} it will be shown that the switch between the two schemes, as implemented in \code, works seamlessly, moving smoothly from $v = 0.563624$ to $u^\tau = 1.21061$ across the switching point. 

When interpolating the EoS table to obtain the derivatives needed on the r.h.s. of Eqs.~(\ref{eq-pvderivative},\ref{eq-puderivative}) one can encounter numerical errors in regions where derivatives of the EoS change discontinuously, e.g. in the recently developed BEST EoS \cite{Parotto:2018pwx} which adds (using a certain prescription) a critical point and first-order phase transition to the LQCD-HRG interpolated EOS4. For such situations, \code\ offers another option for the root finding that avoids calculating these derivatives, at the price of somewhat degraded convergence which can slow the root finding algorithm compared to the Newton-Raphson method. This second regulation scheme may also be preferred when evolving more than one conserved charge, in which case not having to compute the thermodynamic derivatives may overcompensate for the slower intrinsic convergence of the root finding algorithm.

The modified root finder employs the following simple iteration scheme: starting with an initial guess $v_i$ for the velocity (e.g. the solution at this grid point from the preceding time step), we determine $\bigl(\ed(v_i), \n(v_i)\bigr)$ from Eqs.~(\ref{root-e},\ref{root-n}) and the EoS $\peq(\ed,\n)$, compute an updated value $v_{i+1}$ of the velocity from
\begin{equation}
    v_{i+1} = \frac{M}{M^\tau +\peq(\ed(v_i),\n(v_i)) + \Pi}\,,
\label{root-v-expression-2}
\end{equation}
and iterate these steps until convergence is reached. For $v \geq 0.563624$ or $u^\tau \geq 1.21061$, one instead updates $u^\tau$ using the equation 
\begin{equation}
  u^\tau_{i+1} = 
  \sqrt{\frac{M^\tau+\peq\bigl(\ed(u^\tau_i),\n(u^\tau_i)\bigr)
        +\Pi}
       {\ed +\peq\bigl(\ed(u^\tau_i),\n(u^\tau_i)\bigr)+\Pi}}
\label{root-u-expression}
\end{equation}
until convergence is reached. 

In principle, the two methods are equivalent and should find the same root, within the prescribed numerical precision. In the Gubser test described in Sec.~\ref{gubsertest} they are indeed shown to yield identical numerical results. In both methods, higher numerical precision of the solution should be demanded when solving for $v$, due to the speed limit $v<1$.

We point out that extra hydrodynamic variables are propagated in the code that are not used in the root finding algorithm, such as $\pi^{xx}$, $\pi^{xy}$, $\pi^{x\eta}$, and $\V^\eta$. In principle, these could be computed from the other components of the shear stress and baryon diffusion current by using the tracelessness of $\pi^{\mu\nu}$ and the orthogonality of $\pi^{\mu\nu}$ and $\V^\mu$ to $u^\mu$. Instead, we propagate all shear stress and baryon diffusion components dynamically and use the tracelessness and orthogonality conditions to check the numerical precision of the code.

\subsection{Regulation scheme}
\label{sec-regul}

The solution of the hydrodynamic equations of motion on discretized grids, the bilinear interpolation of the EoS table at each grid point, the iterative nature of the root finding algorithm, and the need for calculating derivatives numerically all engender unavoidable numerical errors. In addition, the numerical solution for the hydrodynamic variables can make excursions into regions where the approximations under which the evolution equations were derived (such as ignoring higher-order gradient terms) are no longer valid, and the numerical evolution algorithm produces unphysical results. This happens, in particular, because Nature provides us with initial conditions that exhibit unavoidable quantum fluctuations which can lead to local excursions outside the region of validity of dissipative hydrodynamics. Although dissipation usually erases such large fluctuations over short time scales \cite{Shen:2014vra, Bazow:2016yra}, this may not happen quickly enough to avoid numerical instability of the evolution algorithm.\footnote{%
    Note that we are not even talking about thermal fluctuations during the hydrodynamic evolution (our code solves deterministic equations of motion) which add possibly large stochastic fluctuations throughout the evolution history \cite{Singh:2018dpk,Sakai:2018sxp}.}

In realistic event-by-event simulations such fluctuations can result in large gradients in both space and time which the code has to be able to deal with. Large gradients of the macroscopic variables can yield locally large Knudsen numbers (for which the fluid dynamic approximation breaks down) or large inverse Reynolds numbers (in which case the applicability of the 14-moment approximation used to simplify the hydrodynamic equations of motion is doubtful) \cite{Shen:2014vra,Bazow:2016yra}. To ensure numerical stability of the code, such excursions must be regulated. To avoid the undesirable consequence that, after regulation, the algorithm no longer solves the underlying evolution equations, the regulation must be local, i.e. it must affect only very localized space-time regions, and its effects must be monitored so that the user is warned when the regulation becomes so strong and the regulated regions become so large that the code no longer correctly simulates the physics encoded in the evolution equations. 

In practice, large gradients can drive large shear stress, bulk viscous pressure and baryon diffusion currents, and these can result in numerical instability or failure of the root finding algorithm. When this happens it is typically during the earliest evolution stage (where both the physical inhomogeneities driven by quantum fluctuation and the longitudinal expansion rate are largest) and/or in the very dilute regions near the transverse and longitudinal edge of the computing grid where the dissipative corrections to the leading thermodynamic quantities become large and the matter can no longer be reasonably treated as a fluid. Since the latter regions are typically far outside the domain where the matter is in the quark-gluon plasma phase (and thus outside the region where we want to apply the hydrodynamic picture), regulating them is innocuous as long as the regulation effects do not have sufficient time to propagate back inwards into the QGP region. Regulating large initial fluctuations is more dangerous because the fluctuations can be large both in- and outside the QGP phase. Both types of regulations must be carefully monitored.

Regulation schemes can be tricky, and a variety of implementations exist.\footnote{%
    For example, CLVisc \cite{Pang:2018zzo} requires $\texttt{max} (|\pi^{\mu\nu}|) < T_0^{\tau\tau}$; when this is violated for some cell in the dilute region, $\pi^{\mu\nu}$ is set to 0 locally. vHLLE \cite{Karpenko:2013wva} requires $\texttt{max}(|\pi^{\mu\nu}|/|T_0^{\mu\nu}|) < C$ and $|\Pi|/\peq < C$, with $C$ being a constant of order but smaller than 1; if one of these conditions is violated, $\Pi$ and/or $\pi^{\mu\nu}$ are rescaled by a factor (which is common for all components of $\pi^{\mu\nu}$) to satisfy this requirement.}
In \code, we follow the lead of {\sc iEBE-VISHNU} \cite{Shen:2014vra}, {\sc GPU-VH} \cite{Bazow:2016yra} and {\sc MUSIC} \cite{Denicol:2018wdp} and implement two types of regulation that build on the schemes suggested in these earlier codes. Both are triggered by large dissipative flows which are then regulated. The trigger criterion compares (in ways defined more precisely below) $\pi^{\mu\nu}$ with $T_0^{\mu\nu}=\ed u^\mu u^\nu - \peq\Delta^{\mu\nu}$, $\Pi$ with $\sqrt{\ed^2+3\peq^2}$, and $\V^\mu$ with $\n u^\mu$. 

\begin{figure}[!hbtp]
    \centering
    \includegraphics[width=0.9\textwidth]{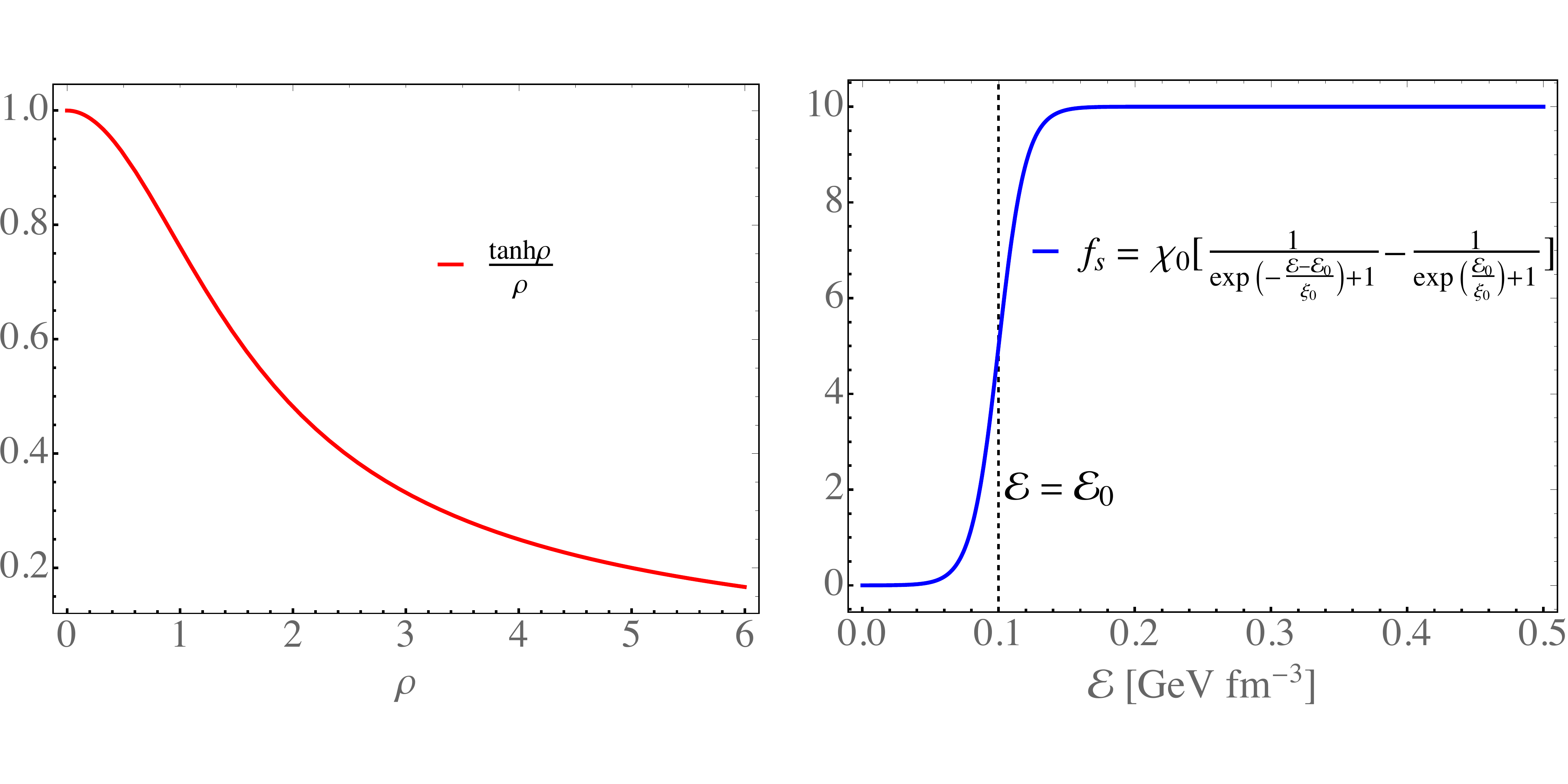}
    \caption{The regulation functions used in our two regulation schemes. {\sl Left:} Regulation function used in the first scheme. Large dissipative components yield large values of $\rho$ where the regulation is stronger. {\sl Right:} Regulation strength function $f_s$ used in the second regulation scheme, for the parameter choice $\chi_0 = 10$, $\ed_0 = 0.1$\,GeV/fm$^3$ and $\xi_0 = 0.01$\,GeV/fm$^3$. $\ed_0$ is the critical energy density below which strong regulation kicks in.
    \label{F3} 
    }
\end{figure}

For the shear stress tensor, {\sc iEBE-VISHNU} \cite{Shen:2014vra} and {\sc GPU-VH} \cite{Bazow:2016yra} require 
\begin{equation}
    \sqrt{\pi^{\mu\nu}\pi_{\mu\nu}} \leq \rho_\mathrm{max}\sqrt{T_0^{\mu\nu}T_{0\mu\nu}} = \rho_\mathrm{max}\sqrt{\ed^2+3\peq^2}\,,
\end{equation}
with $\rho_\mathrm{max}\leq 1$. In addition, the tracelessness of $\pi^{\mu\nu}$ and its orthogonality to $u_\nu$ are required,
\begin{equation}
    \pi^{\mu}_{\mu}\leq\xi_0\sqrt{\pi^{\mu\nu}\pi_{\mu\nu}}\quad
    \textrm{and}\quad \pi^{\mu\nu}u_\nu\leq\xi_0\sqrt{\pi^{\mu\nu}\pi_{\mu\nu}}\;,
\end{equation}
where $\sqrt{\pi^{\mu\nu}\pi_{\mu\nu}}$ sets the scale and $\xi_0\ll 1$ is a small number \cite{Shen:2014vra}. At grid points where these trigger conditions are violated, Refs.~\cite{Shen:2014vra, Bazow:2016yra} regulate the shear stress tensor $\pi^{\mu\nu}$ by (see the left plot of Fig.~\ref{F3})
\begin{equation}
  \pi^{\mu\nu}\to\frac{\tanh\rho_\pi}{\rho_\pi}\pi^{\mu\nu}\;,
\label{regPimunu}
\end{equation}
where
\begin{equation}
  \rho_\pi\equiv\texttt{max}
  \left[
  \frac{\sqrt{\pi^{\mu\nu}\pi_{\mu\nu}}}
       {\rho_\mathrm{max}\sqrt{\ed^2+3\peq^2}}\,,\quad
  \frac{g_{\mu\nu}\pi^{\mu\nu}}
       {\xi_{0}\rho_\mathrm{max}\sqrt{\pi^{\mu\nu}\pi_{\mu\nu}}}\,,\quad
  \frac{\pi^{\lambda\mu}u_{\mu}}
       {\xi_{0}\rho_\mathrm{max}\sqrt{\pi^{\mu\nu}\pi_{\mu\nu}}}\; \forall\;\lambda
  \right]. 
\label{rhoPimunu}
\end{equation}

For the bulk viscous pressure, which can make the root finding process fail when negative and too large, Refs. \cite{Shen:2014vra, Bazow:2016yra} regulate $\Pi$ during the root finding process to ensure existence of at least one non-negative solution for $\ed$, $v$ or $u^\tau$. In \code\ we demand instead that 
\begin{equation}
  \sqrt{3\Pi^2} \leq \rho_\mathrm{max}\sqrt{\ed^2+3\peq^2}\;,
\label{Pilimit}
\end{equation}
and where this trigger condition is violated we regulate $\Pi$ by
\begin{equation}
  \Pi\to\frac{\tanh\rho_\Pi}{\rho_\Pi}\Pi\;,
\label{regPi}
\end{equation}
with
\begin{equation}
\rho_\Pi\equiv \frac{1}{\rho_\mathrm{max}}\sqrt{\frac{3\Pi^2}{\ed^2+3\peq^2}}\;.
\label{rhoPi}
\end{equation}

In the same spirit, we require for the (space-like) baryon diffusion current
\begin{equation}
    -\V^\mu \V_\mu \ll \n^2 \quad\textrm{and}\quad \V^\mu u_\mu = 0\;.
\end{equation}
In the code we replace these conditions by
\begin{equation}
    \sqrt{-\V^\mu \V_\mu} \leq \rho_\mathrm{max}\sqrt{\n^2} \quad\textrm{and}\quad \V^\mu u_\mu \leq \xi_0\sqrt{-\V^\mu \V_\mu}\;.
\end{equation}
When one of these conditions is violated in a cell it triggers the following regulation of the baryon diffusion current:
\begin{equation}
  \V^\mu\to\frac{\tanh\rho_\V}{\rho_\V}\V^\mu\;,
\label{regV}
\end{equation}
with
\begin{equation}
  \rho_\V\equiv\texttt{max}
  \left[
  \frac{\sqrt{-\V^\mu \V_\mu}}{\rho_\mathrm{max}\sqrt{\n^2}}\,,\quad
  \frac{\V^\mu u_\mu}{\xi_0\rho_\mathrm{max}\sqrt{-\V^\mu \V_\mu}}
  \right]\;.
\label{rhoV}
\end{equation}
%

Equations (\ref{regPimunu},\ref{regPi},\ref{regV}) define the first of our two regulation schemes. In contrast to Refs.~\cite{Shen:2014vra, Bazow:2016yra} where $\Pi$ was regulated during the root finding process, our regulation here is performed only after finishing each step of the two-step RK-KT algorithm. After implementing it we find that there is no need for additional regulation of $\Pi$ during the root finding process. As defaults for the regulation parameters we use the same values $\xi_0 = 0.1$ and $\rho_\mathrm{max} = 1$ as proposed in \cite{Shen:2014vra, Bazow:2016yra}. For documentation of the sensitivity studies leading to these default values we refer the reader to Ref.~\cite{Shen:2014vra}.

A second, different regulation scheme is based on the one implemented in the latest version of {\sc MUSIC} \cite{Denicol:2018wdp}. (A flag in the code allows the user to select the preferred regulation scheme before running it.) In this second scheme, the dissipative components are rescaled by
\begin{equation}
    \pi^{\mu\nu}\to\frac{r^\mathrm{max}_\pi}{r_\pi}\pi^{\mu\nu}\;,\quad \Pi\to\frac{r^\mathrm{max}_\Pi}{r_\Pi}\Pi\;,\quad  \V^\mu\to\frac{r^\mathrm{max}_\V}{r_\V}\V^\mu\;,\label{eq-rmaxr}
\end{equation}
where regulation of quantity $i$ is triggered whenever $r_i$ exceeds the corresponding maximally allowed value $r_i^\mathrm{max}$ $(i=\pi, \Pi, \V)$,\footnote{%
    If the regulation for the shear stress or baryon diffusion is triggered, all components of $\pi^{\mu\nu}$ or $n^\mu$ are regulated by a common regulation factor.}
with $r_i$ defined by 
\begin{equation}
    r_\pi = \frac{1}{f_s}\sqrt{\frac{\pi^{\mu\nu}\pi_{\mu\nu}}{\ed^2+3\peq^2}}\;,\quad 
    r_\Pi = \frac{1}{f_s}\sqrt{\frac{3\Pi^2}{\ed^2+3\peq^2}}\;,\quad 
    r_\V = \frac{1}{f_s}\sqrt{\frac{-\V^\mu \V_\mu}{\n^2}}\;.
\label{eq-r}
\end{equation}
By comparing Eq.~(\ref{eq-rmaxr}) to Eqs.~(\ref{regPimunu},\ref{regPi},\ref{regV}) and Eq.~(\ref{eq-r}) to Eqs.~(\ref{rhoPimunu},\ref{rhoPi},\ref{rhoV}) above, we see that the factors $r_i^\mathrm{max}/r_i$ play the same role as $\tanh{\rho_i}/\rho_i$ in the first regulation scheme, causing stronger regulation for larger $r_i$, with $r_i$ playing the role of the quantity $\rho_i$ whereas $f_s$, defined by
\begin{equation}
    f_s = \chi_0
    \left[\frac{1}
               {\exp\bigl(-(\ed{-}\ed_0)/\xi_0\bigr)+1}
         -\frac{1}{\mathrm{exp}(\ed_0/\xi_0)+1}
    \right],
\end{equation}
playing a similar role as $\rho_\mathrm{max}$: as $f_s$ or $\rho_\mathrm{max}$ grows larger, the regulation gets weaker. $f_s$ is designed to approach $\chi_0$ when $\ed\gg\ed_0$ and 0 when $\ed\ll\ed_0$.

The right plot of Fig.~\ref{F3} shows that for $\ed < \ed_0$, i.e. in the dilute region, $f_s$ decreases exponentially and the regulation strength increases accordingly. On the other hand, for large values of the parameter $\chi_0$, the regulation will hardly ever be triggered in the dense region $\ed>\ed_0$. Thus, unlike the first method, which always regulates larger dissipative components more strongly, irrespective of the energy density at the grid point, for the choice $\ed_0=0.1$\,GeV/fm$^3$ \cite{Denicol:2018wdp} the second method causes hardly any regulation at grid points in the dense QGP region but more frequent and stronger regulation in the dilute region far outside the QGP fluid. The authors of Ref.~\cite{Denicol:2018wdp} used $\chi_0=10$, and chose $r^\mathrm{max}_\V = 1$ for regulating the baryon diffusion current; we adopt the same value for $\chi_0$ and identical maximum $r$ values for all dissipative flows: $r^\mathrm{max}_\pi = r^\mathrm{max}_\Pi = r^\mathrm{max}_n = 1$. 

The default set for the regulation parameters is not universal and may need adjustments for different initial conditions, collision systems, and collision energies. The user is encouraged to play with these parameters to achieve maximal code stability with minimal changes to the physics encoded in the evolution equations. The regulation scheme may need to become more involved in future versions of dissipative hydrodynamics that include possibly large thermal and/or critical fluctuations in the dynamics (see e.g. \cite{Singh:2018dpk, Sakai:2018sxp}).

\section{Code validation with semi-analytical solutions}
\label{numericaltests}

Numerical codes solving second-order (``causal'') relativistic dissipative fluid dynamics in 3+1 dimensions have only been developed over the last decade. They solve a problem for which in general no analytic solutions are available. Careful validation of any such code by testing its various components in simplified settings for which analytic or semi-analytic solutions {\it are} available is therefore mandatory. Some of these tests are nowadays standard and are included with this distribution precisely for that reason. The precursor of this code, the CPU-version of {\sc GPU-VH} \cite{Bazow:2016yra} was carefully validated using similar tests, but the entire baryon evolution sector in \code\ is new such that direct comparisons with {\sc GPU-VH} are of limited value. We therefore include here in particular novel semi-analytic tests of the baryon evolution equations. Direct code-to-code comparisons with {\sc MUSIC} (whose latest version \cite{Denicol:2018wdp} also includes baryon evolution) will become possible when that version of {\sc MUSIC} becomes public.

Building on validation protocols described in Refs.~\cite{Karpenko:2013wva, Shen:2014vra, Bazow:2016yra, Pang:2018zzo, Denicol:2018wdp}, we here discuss tests in which we compare, for identical initial conditions, the numerical solutions from \code\ with (semi-)analytic solutions using {\sc Mathematica} \cite{Mathematica} for the Riemann problem for the Euler equations \cite{toro2009riemann, sod1978survey, marti_muller_1994, RISCHKE1995346, RISCHKE1995383}, for Bjorken flow \cite{PhysRevD.27.140}, and for Gubser flow \cite{PhysRevD.82.085027,Gubser:2010ui} extended to systems with non-zero net baryon density and baryon diffusion current induced by a fluctuation in the initial state. We also include a direct comparison of \code\ with the independently developed numerical algorithm described in Ref.~\cite{PhysRevC.86.014908} for a system with non-zero net baryon density in a (1+1)-dimensional setting with general longitudinal but vanishing transverse flow. By generalizing previously developed validation protocols to systems with non-zero net baryon density and baryon diffusion currents the work described in this section prepares the ground for code validation of other hydrodynamic codes at finite baryon density that are presently being developed elsewhere for the study of heavy-ion collisions at BES energies.

All tests described in this section are done without code regulation, i.e. all the regulation schemes described in Sec.~\ref{sec-regul} are turned off. In the code, all dimensionful quantities are represented by numbers given in length units, using the appropriate powers of [fm]; when plotting the results we sometimes convert them to physical units by multiplying with the appropriate powers of $\hbar c = 0.197$\,GeV\,fm.

In the tests and all other applications of the code completed to date we have used the following grid spacings: $\Delta x = \Delta y = 0.05$~fm and $\Delta\eta_s = 0.02$; $\Delta\tau$ is adjusted as needed and described in each case below. In realistic simulations the choice of grid spacing has to be a compromise between computational economy and capturing relevant physical information (e.g. large gradients, especially at early times). For smooth, ensemble-averaged initial conditions, larger grid spacings than those listed above can be used, whereas simulating small collision systems (such as proton-proton collisions) may require finer grids.

\subsection{The Riemann problem}
\label{sec4.1}

\begin{figure}[!htbp]
    \centering \includegraphics[width=\textwidth]{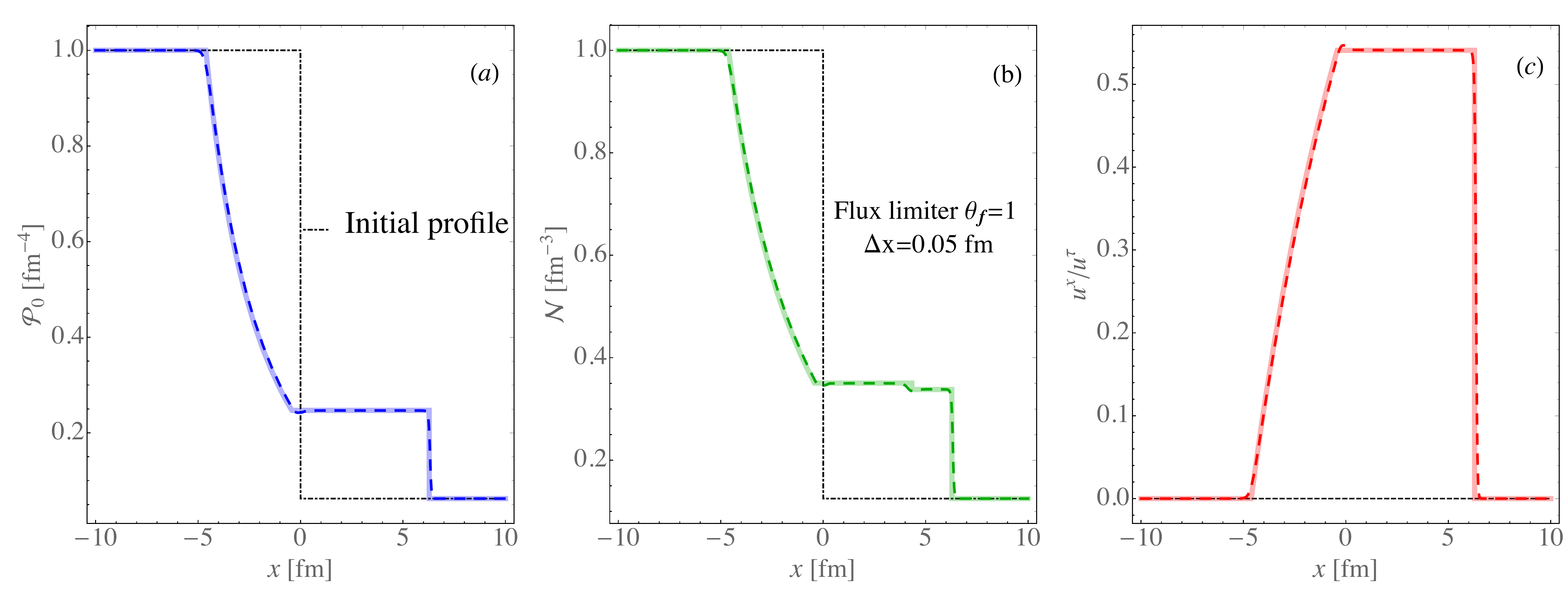}
    \caption{Analytical (Ref.~\cite{PhysRevC.98.035201}, continuous lines) vs. numerical results (\code, broken lines) for Sod's relativistic shock-tube problem in an ideal fluid with a conformal EoS: (a) pressure; (b) baryon density; (c) scaled velocity $u^x/u^\tau$. The numerical simulation starts at $\tau_0 = 0.5$\,fm/$c$, the grid spacing in the transverse $(x,y)$ plane is $\Delta x=\Delta y = 0.05$\,fm, and the flux limiter is set to $\theta_\mathrm{f} = 1$. The initial conditions to the left ($l$) and right ($r$) of the shock discontinuity are $\p_l{\,=\,} 1$\,fm$^{-4}$,\; $\n_l{\,=\,}1$\,fm$^{-3}$,\; $u^x_l{\,=\,}u^x_r{\,=\,}0$,\; $\p_r{\,=\,}0.0625$\,fm$^{-4}$,\; and $\n_r{\,=\,}0.125$\,fm$^{-3}$. The plots show results at $\tau=8.5$\,fm/$c$ for the numerical and $\tau=8.0$\,fm/$c$ for the analytical solution (see text for explanation).
    \label{F4}
    }
\end{figure}

We start with testing the code against an analytical solution of the Riemann problem for the Euler equations, which historically has played an important role in fluid dynamics research and in the development of hydrodynamic codes \cite{toro2009riemann, marti_muller_1994}. Here a special case of the relativistic Riemann problem, known as Sod's shock tube problem \cite{sod1978survey, marti_muller_1994, RISCHKE1995346, RISCHKE1995383}, is considered describing the 1D evolution of an ideal fluid in the transverse plane during the decay of a discontinuity across a $(y,z)$ surface placed at $x{\,=\,}0$, separating two constant initial states (``left'' ($l$) for $x<0$ and ``right'' ($r$) for $x>0$) at rest:
\begin{equation}
    (\p,\n,u^x) =
    \begin{cases}
         (\p_l, \n_l, u^x_l = 0)\;,\qquad x < 0\;,\\
         (\p_r, \n_r, u^x_r = 0)\;,\qquad x > 0\;.
    \end{cases}
\label{eq-riemann-ic}
\end{equation}
In the longitudinal ($z$) direction the fluid is assumed to expand with a boost-invariant velocity profile $u^z/u^\tau{\,=\,}z/t$ (i.e. $u^\eta{\,=\,}0$), and the system is initialized along a surface of constant longitudinal proper time $\tau_0$.

The decay of this initial discontinuity gives rise to general features of the Riemann problem, characterized by three elementary waves. Two of them are a rarefaction wave and a shock, moving into the two initial state regions of high and low density, respectively. Between them, two additional states emerge, separated by the third wave, which is a contact discontinuity moving with the fluid \cite{toro2009riemann, marti_muller_1994} (see Fig.~\ref{F4}(b)). For a conformal EoS, an analytical solution for this problem can be derived from the conservation laws with the boundary condition that across the contact discontinuity pressure and velocity are constant (Figs.~\ref{F4}(a,c)) while the density has a jump (Fig.~\ref{F4}(b)) \cite{marti_muller_1994, Molnar:2009tx, PhysRevC.98.035201}. For non-conformal equations of state at non-zero net baryon density no general analytic solution is known, and the numerical solution can give rise to complex features (see, e.g., \cite{marti_muller_1994, SCHNEIDER199392}). Since for EOS1 $\partial \peq/\partial \n$ (which is needed in Eqs.~(\ref{eq-pvderivative},\ref{eq-puderivative}) for the velocity finding algorithm) is zero, the evolution of baryon density decouples from that of the energy density and pressure. 

In the code, the assumption of longitudinal boost-invariance is implemented by setting the number of cells in the longitudinal direction to 1. The initial profiles of $\ed$ and $T$ are obtained from the EoS. For ideal fluid dynamics the equations of motion become
\begin{align}
    D\n =& -\n\, \theta\;, \label{eq-riemann-n}\\
    D\ed =& -(\ed+\p)\theta\;,\label{eq-riemann-p}\\
    (\ed+\p)Du^\mu =& - \Delta^{\mu \nu} \partial_\nu \p\;. \label{eq-riemann-eom}
\end{align}
They are solved with initial conditions (\ref{eq-riemann-ic}) with the default parameters listed in Fig.~\ref{F4}. For details about the analytical solution we refer the reader to Ref.~\cite{PhysRevC.98.035201} (see also \cite{RISCHKE1995346, RISCHKE1995383}). We point out that in the analytical solution from Ref.~\cite{PhysRevC.98.035201} the evolution starts at time zero whereas in the code the hydrodynamic evolution is initialized at $\tau_0 = 0.5$\,fm. Since the solution is self-similar and depends only on the variable $x/(\tau-\tau_0)$, we therefore compare in Fig.~\ref{F4} the numerical results at $\tau=8.5$\,fm/$c$ to the analytical solution at $\tau-\tau_0=8.0$\,fm/$c$. For simplicity, the numerical test is done in Cartesian coordinates where all Christoffel symbols vanish. Also, to adequately capture the large discontinuity in the initial state, derivatives should be evaluated using Eq.~(\ref{eq-cendiff}), and not Eq.~(\ref{eq-approx-derivative}) which would yield zero initial gradients and result in no evolution at all.

Figure~\ref{F4} demonstrates very good overall agreement between the analytical and numerical solutions; the shocks and contact discontinuities are well captured. Although the baryon evolution is decoupled, this test still demonstrates excellent performance of the root finding algorithm. 

\subsection{Bjorken flow}
\label{sec4.2}

In this subsection we test \code\ in Milne coordinates for a transversally homogeneous dissipative fluid undergoing longitudinally boost-invariant Bjorken expansion \cite{PhysRevD.27.140}. Boost-invariant systems are characterized by space-time rapidity independent macroscopic observables and a flow profile that looks static (i.e. $u^\mu = (1,0,0,0)$) in Milne coordinates \cite{Jeon:2015dfa}. In spite of experimental evidence for longitudinal density gradients, there are strong phenomenological indications that near mid-rapidity a longitudinally boost-invariant flow profile is a good approximation for relativistic heavy-ion collisions at $\sqrt{s}\gtrsim 100$\,GeV/nucleon (see, e.g., \cite{Jeon:2015dfa}). The additional assumption of transverse homogeneity, however, is clearly unrealistic, given the finite transverse size of the colliding nuclei. Still, it provides a useful test bed because the resulting independence of all macroscopic quantities from all three spatial dimensions simplifies the dissipative hydrodynamic evolution equations to a set of coupled ordinary differential equations which can be solved with {\sc Mathematica}:
\begin{eqnarray}
\dot{\ed} &=&-\frac{\ed+\peq+\Pi -\pi }{\tau }\;,
\label{evolutionEd} 
\\
\tau _{\Pi }\dot{\Pi}+\Pi &=&-\frac{\zeta }{\tau }-\delta _{\Pi \Pi }\frac{%
\Pi }{\tau }+\lambda _{\Pi \pi }\frac{\pi }{\tau }\;,
\label{evolutionPi} 
\\
\tau _{\pi }\dot{\pi}+\pi &=&\frac{4}{3}\frac{\eta}{\tau} -\left( \frac{1}{3}\tau
_{\pi \pi }+\delta _{\pi \pi }\right) \frac{\pi }{\tau }+\frac{2}{3}\lambda _{\pi \Pi }%
 \frac{\Pi}{\tau}\;,
\label{evolutionpi}
\\
%
\dot{\n} &=&-\frac{\n}{\tau }\;,
\label{evolutionn}
\\
\tau_{\V} \dot{\V}^\eta+\V^\eta&=&-\left(\tau_{\V}+\delta_{\V\V}+\frac{2}{3}\lambda_{\V\V}\right)\frac{\V^\eta}{\tau}\;,
\label{evolutionneta}
\end{eqnarray}
where $\pi\equiv-\tau^2\pi^{\eta\eta}$ has been introduced. 

\begin{figure*}[!htbp]
    \centering
    \includegraphics[width= 0.95\textwidth]{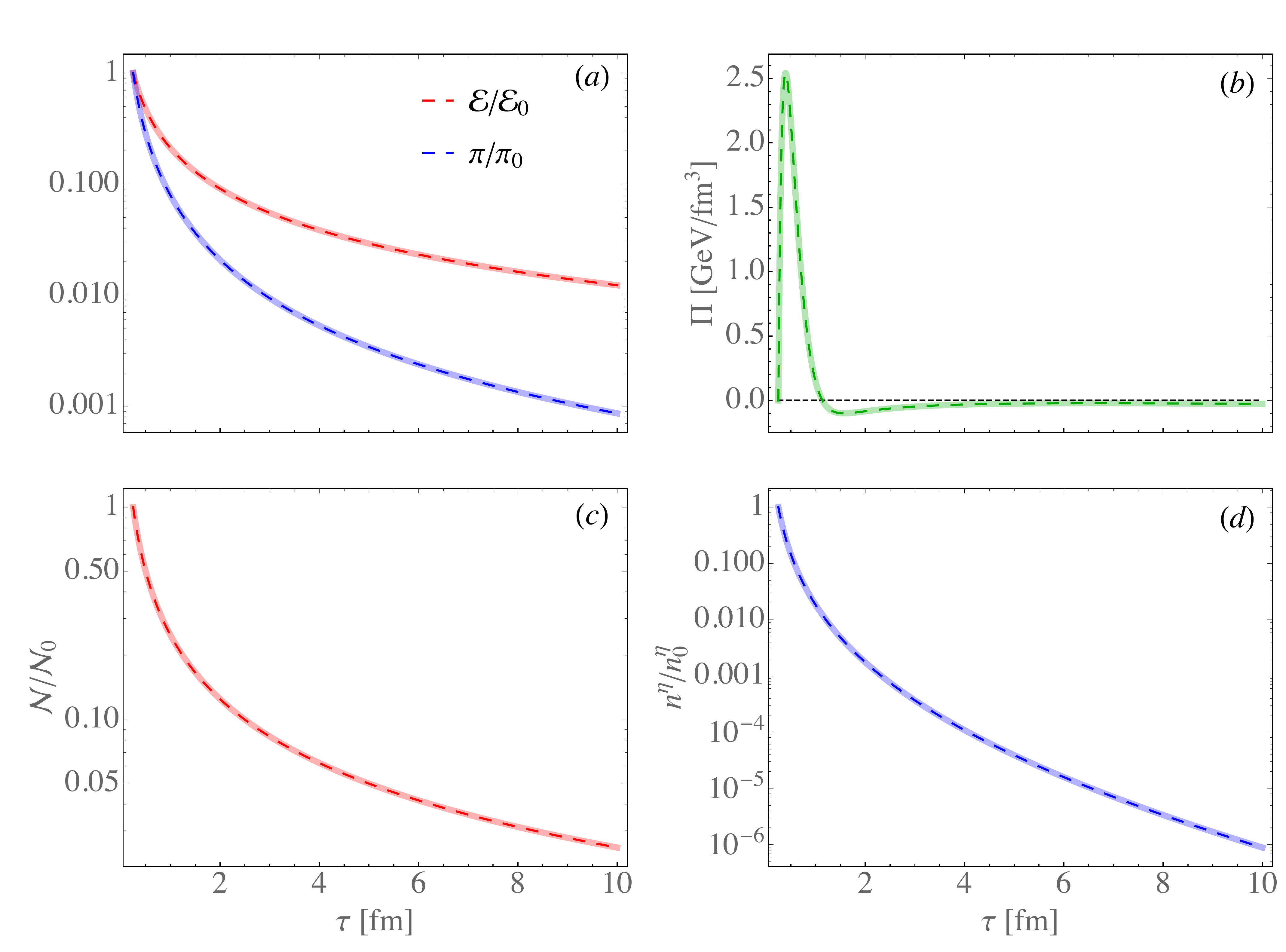}
    \caption{Comparing the semi-analytic results (continuous lines) with numerical output from \code\ (broken lines) for a fluid undergoing Bjorken expansion under the influence of the  Wuppertal-Budapest EoS (EOS2): (a) normalized energy density $\ed$ and shear stress $\pi=-\tau^2\pi^{\eta\eta}$; (b) bulk viscous pressure $\Pi$; (c) normalized baryon density $\n$; (d) normalized baryon diffusion current $n^\eta$. We use $\theta_\mathrm{f} = 1$ for the flux limiter, $\bar\eta=0.2$ for the kinematic shear viscosity, and the parametrization (\ref{eq-zetas}) for the kinematic bulk viscosity $\bar\zeta$. From these the relaxation times $\tau_\pi$, $\tau_\Pi$ and the viscosity related transport coefficients are computed using Eqs.~(\ref{etatau}-\ref{eq-zetas}). For $\tau_n$ we use Eq.~(\ref{taun}) with $C_B=4.0$ when $c\tau_n$ and $1/T$ are measured in fm, and Eq.~(\ref{diffcoeff}) for the baryon related second-order transport coefficients. The expansion is initialized at $\tau_0 = 0.25$\,fm/$c$ with initial conditions $T_0 = 4.5$\,fm$^{-1} = 0.89$\,GeV, 
    $\n_0= 500$ fm$^{-3}$, $\Pi_0=0$, $\pi_0 = \pi_{\mathrm{NS}} = \frac{4}{3} \frac{\eta}{\tau_0}$, and $n_0^\eta=10$ fm$^{-4}$. The initial energy density and pressure are obtained from the EoS.
    \label{F5}
    }
\end{figure*}

In the \code\ simulation the numbers of cells in $(x, y, \eta_s)$ directions are all set to 1.\footnote{%
    Of course, one can also set the number of cells in $(x, y)$ larger than 1 but to ensure transverse spatial homogeneity quantities like $\ed$ and $\peq$ should be the same in all cells.}
With boost-invariant and transversally homogeneous initial conditions, the numerical results from the (3+1)D \code\ code can be tested against a solution of the coupled ODEs (\ref{evolutionEd}-\ref{evolutionneta}) from a separate {\sc Mathematica} code. As for the Riemann problem, the baryon evolution decouples from the rest of the system if an EoS $\peq = \peq(\ed)$ is used; in this case, even with a nonzero baryon diffusion current in longitudinal direction, the Bjorken expansion remains unchanged.
Although longitudinal boost-invariance and transverse homogeneity do not allow any gradients of the chemical potential that could drive a baryon diffusion current, we can then still test the baryonic sector of the code by initiating it with nonzero initial values for the net baryon density and diffusion current. These out-of-equilibrium initial values will then relax according to Eqs.~(\ref{evolutionn},\ref{evolutionneta}), without affecting the Bjorken flow profile.  

By using EOS2 (the Wuppertal-Budapest EoS at $\mu=0$) which features a non-zero interaction measure, we can also test the evolution of the bulk viscous pressure $\Pi$ which is propagated by Eq.~(\ref{evolutionPi}) as an additional dissipative degree of freedom \cite{Bazow:2016yra}. Noting that Eq.~(\ref{evolutionPi})
involves a rather complex parametrization of the transport coefficients (\ref{betaPi})-(\ref{eq-zetas}), we emphasize that this test is indeed non-trivial.

As Fig.~\ref{F5} shows, the agreement of \code\ with the semi-analytic {\sc Mathematica} solution is excellent. The evolution of the baryon diffusion current follows the exact decay very precisely over six orders of magnitude. This would not be possible if the root finding algorithm did not perform with high accuracy.\footnote{%
    As for the Riemann problem, this test again checks this property only for the case $\partial\peq/\partial\n = 0$.}

Still, because of boost-invariance and transverse homogeneity many terms in the full set of evolution equations vanish in this example. The absence of any kind of transverse expansion in this test is particularly worrisome. This question will be addressed next with the ``Gubser test''.

%
\subsection{Gubser flow}
\label{gubsertest}
%

Gubser flow \cite{PhysRevD.82.085027,Gubser:2010ui} describes conformally symmetric systems that, in addition to longitudinally boost-invariant Bjorken flow, undergo at the same time strong azimuthally symmetric (``radial'') transverse flow. Contrary to heavy-ion collisions where transverse flow is initially zero and only generated after the collision in response to transverse pressure gradient, in Gubser flow the transverse flow exists at all times (i.e. even on a hypersurface corresponding to some very early ``initial'' longitudinal proper time $\tau_0$). 

\subsubsection{Gubser coordinates}
\label{sec4.3.1}

Gubser flow originates from an ingenious symmetry that, like Bjorken symmetry, makes the flow appear static in an appropriately chosen set of coordinates called ``Gubser coordinates'' which map Minkowski space onto a 3-dimensional de Sitter space times a line, dS$_3\otimes\mathbb{R}$ \cite{PhysRevD.82.085027,Gubser:2010ui}. As a result of this symmetry, macroscopic quantities do not depend on any of the space-like coordinates but only on the time-like coordinate in this system, and the dissipative hydrodynamic equations again reduce to a set of coupled ODEs in that time coordinate, $\rho\in\mathbb{R}$.

To introduce the Gubser coordinates we first rescale the invariant distance measure of Minkowski space in Milne coordinates with a Weyl transformation\footnote{%
    To make our equations readily comparable with those in the by now vast literature on Gubser flow we temporarily (i.e. in this subsection only) switch our metric signature convention to the mostly-plus metric, i.e. $g^{\mu\nu} = \textrm{diag}(-1,1,1,1)$ in Cartesian coordinates.}  
\begin{equation}
    ds^2\to d\hat s^2\equiv ds^2/\tau^2 = (-d\tau^2+dr^2+r^2d\phi^2)/\tau^2+d\eta_s^2\;.
    \label{eq-weyl}
\end{equation}
Next we perform the coordinate transformation $x^\mu = (\tau, r, \phi, \eta_s) \to \hat x^\mu = (\rho,\theta,\phi, \eta_s)$,\footnote{%
    Here $r\equiv\sqrt{x^2+y^2}$ and $\phi\equiv\tan^{-1}(x/y)$. All quantities expressed as functions of Gubser coordinates are made unitless by scaling them with the appropriate powers of the Milne time $\tau$ and labeled with a hat.} 
by introducing \cite{PhysRevD.82.085027,Gubser:2010ui}
\begin{align}
    \rho(\tau,r) &\equiv - \sinh^{-1}\left(\frac{1-q^2\tau^2+q^2r^2}{2q\tau}\right)\;,\label{eq-rho}\\
    \theta(\tau,r) &\equiv\tanh^{-1}\left(\frac{2qr}{1+q^2\tau^2-q^2r^2}\right)\;,\label{eq-theta}
\end{align}
where $q$ is an arbitrary energy scale that defines the physical size of the system (the solution is invariant under a common rescaling of $q$, $\tau$ and $r$ such that $qr$ and $q\tau$ remain unchanged). In these coordinates the Weyl-rescaled invariant distance measure becomes
\begin{equation}
    d\hat s^2 = -d\rho^2+\cosh^2\rho\,
    \bigl(d\theta^2+\sin^2\theta d\phi^2\bigr)+d\eta_s^2\;,
\label{eq-measure-desitter}
\end{equation}
with the metric 
\begin{equation}
    \hat g_{\mu\nu} = \textrm{diag}(-1,\, \cosh^2\rho,\, \cosh^2\rho\sin^2\theta,\, 1)\;.
\label{eq-metric-desitter}
\end{equation}
A system that appears static in the coordinates $\hat x^\mu = (\rho, \theta, \phi, \eta_s)$, i.e. has flow velocity $\hat u^\mu=(1,0,0,0)$, is said to exhibit Gubser flow in Minkowski space. 

To map quantities expressed in Gubser coordinates back to Milne coordinates in Minkowski space one uses metric rescaling \cite{Gubser:2010ui} and the definitions (\ref{eq-rho},\ref{eq-theta}), for example
\begin{align}
    u_{\mu}(\tau, r) &=\tau\frac{\partial \hat x^\nu}{\partial x^\mu}\hat u_\nu(\rho(\tau, r))\;,
\label{eq-gubser-umu}
\\
    \pi_{\mu\nu}(\tau, r) &=\frac{1}{\tau^2}\frac{\partial \hat x^\alpha}{\partial x^\mu}\frac{\partial \hat x^\beta}{\partial x^\nu}\hat\pi_{\alpha\beta}(\rho(\tau, r))\;,
\\
T(\tau, r) &= \hat T(\rho(\tau, r))/\tau \;,
\\
\ed(\tau, r) &= \hat \ed(\rho(\tau, r))/\tau^4 \;.
\label{eq-gubser-energy}
\end{align}
With these transformation rules the Gubser flow profile can be expressed in Milne coordinates through the components
\begin{align}
u^\tau(\tau, r) &= \cosh\kappa(\tau,r)\;, 
\label{eq-gubser-ut}
\\
u^{x}(\tau, r) &= \frac{x}{r}\sinh\kappa(\tau,r)\;,
\label{eq-gubser-ux}
\\
u^{y}(\tau, r) &= \frac{y}{r}\sinh\kappa(\tau,r)\;,
\label{eq-gubser-uy}
\\
u^\phi(\tau, r) &= u^\eta(\tau, r) = 0\;,
\label{eq-gubser-uphi}
\end{align}
where $\kappa(\tau,r)$ is the transverse flow rapidity, corresponding to the transverse flow velocity 
\begin{equation}
    v_\perp(\tau,r)=\tanh \kappa(\tau,r) \equiv 
    \frac{2q^2\tau r}{1+q^2\tau^2+q^2r^2}\;.
\label{eq-gubser-kappa}
\end{equation}
Note the transverse flow components are azimuthally symmetric. This flow is dictated by symmetry so it applies to both ideal and dissipative fluids whose thermodynamic functions have Gubser symmetry (i.e. depend only on $\rho$ when expressed in Gubser coordinates). Different initial conditions for the hydrodynamic quantities and different transport coefficients yield different $\rho$ dependencies for their evolution, translating into different characteristics $r(\tau)$ and different $(\tau, r(\tau))$ profiles when expressed in Minkowski space coordinates. 

To ensure invariance of the hydrodynamic equations under the Weyl transformation (\ref{eq-weyl}) the energy momentum tensor must be traceless. This means that a conformal EoS must be used and the  bulk viscosity and bulk viscous pressure must be set identically to zero.
 
\subsubsection{Gubser flow with baryon diffusion}
\label{sec4.3.2}

In this work we extend the existing semi-analytical solutions for conformal Israel-Stewart hydrodynamics with Gubser flow \cite{PhysRevC.91.014903} to systems with a non-zero baryon diffusion current. One could argue that the longitudinal reflection symmetry under $\eta_s\to-\eta_s$ in the Gubser symmetry indicates that the baryon diffusion current should be zero \cite{Denicol:2018wdp}. However, when the conformal EoS is used, as required by the Weyl invariance of the hydrodynamics, the baryon evolution decouples from the rest of the system, and the baryon density and diffusion current evolve as background fields. This means that a non-zero baryon diffusion current does not modify the Gubser flow profile, and the numerical results for the evolution of the baryon diffusion current can be tested by comparing them to the semi-analytical solutions of the equations of motion obtained from $\hat u^\mu=(1,0,0,0)$ under Gubser symmetry. 

\begin{figure*}[!htbp]
    \centering
    \includegraphics[width= 0.90\textwidth]{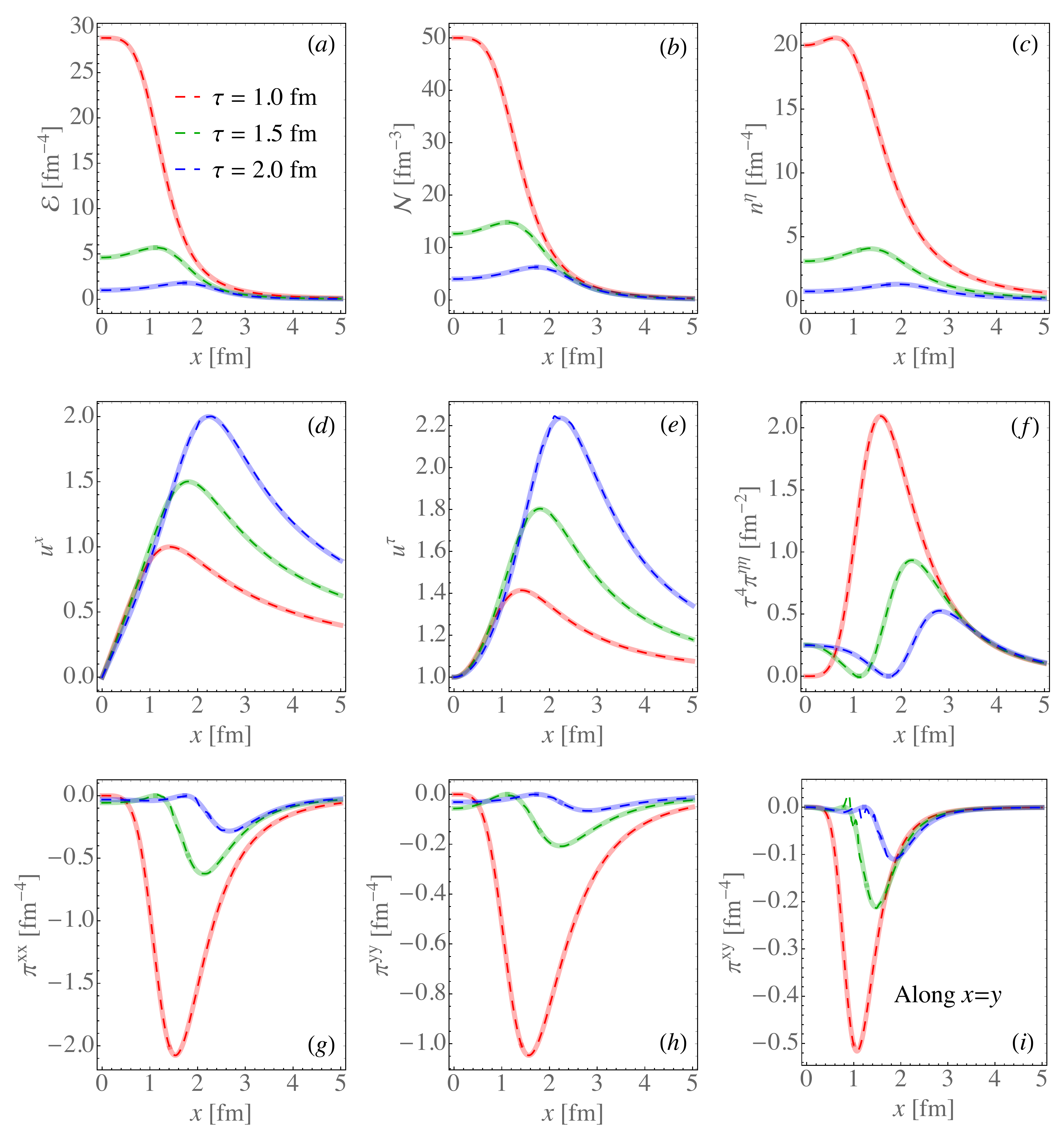}
    \caption{Comparison for Gubser flow between the semi-analytical solutions (continuous lines) and the numerical results from \code\ (using the Newton–Raphson method as root-finder) (broken lines) at $\tau = 1.0,\; 1.5,\; 2.0$ fm$/c$: (a) energy density $\ed$; (b) net baryon density $\n$; (c) baryon diffusion current $n^\eta$; (d),(e) flow velocity components $u^x$ and $u^\tau$; (f)-(i) shear stress tensor components $\tau^4\pi^{\eta\eta}$, $\pi^{xx}$, $\pi^{yy}$, and $\pi^{xy}$. Except for $\pi^{xy}$ in the last panel, which is plotted along the line $x=y$ (or $\phi=\pi/4$), all quantities are shown as functions of $x$ at $\eta_s=\phi=0$. The comparison is made for $q=1$\,fm$^{-1}$ and $\theta_\mathrm{f} = 1.8$, using temporal and spatial grid spacings $\Delta\tau=0.005$\,fm/$c$ and $\Delta x = \Delta y = 0.05$\,fm. For the transport parameters we use $\bar\eta = 0.2$, $C_B=4$, and $\tau_{\pi\pi} = 0$; the remaining transport coefficients are specified in the text. The simulation starts at $\tau_0 = 1$\,fm/$c$ with the following initial conditions: at $\rho=0$ (or equivalently at $(\tau = \tau_0,\; r=0)$) we set $T = 1.2$\,fm$^{-1}$, $\pi^{\eta\eta}=0$, $\n=50$\,fm$^{-3}$, and $n^\eta=20$\,fm$^{-3}$.
    \label{F6}
    }
\end{figure*}

In this subsection we derive these baryon equations of motion in de Sitter space. We rewrite Eqs.~(\ref{eq:vhydro-N}, \ref{eq:vhydro-E}) and (\ref{eq-n-simple}, \ref{eq-pi-simple}) with the mostly-plus metric tensor and apply the Gubser flow profile $\hat u^\mu=(1,0,0,0)$ and the de Sitter metric (\ref{eq-metric-desitter}) to obtain\footnote{%
    Similar to Bjorken flow, the shear stress for Gubser flow has only one independent component for which we take $\hat\pi^{\eta\eta}$. The other non-vanishing components $\hat\pi^{\theta\theta}$ and $\hat\pi^{\phi\phi}$ are related to $\hat\pi^{\eta\eta}$ by tracelessness (which gives $\hat\pi^\eta_\eta = -\hat\pi^\theta_\theta - \hat\pi^\phi_\phi$) and azimuthal symmetry (which implies $\hat\pi^\eta_\eta = -2\hat\pi^\theta_\theta = -2\hat\pi^\phi_\phi$ for evolution with azimuthally symmetric initial conditions).
    }
\begin{eqnarray}
   \partial_\rho\hat \ed + 2\tanh\rho\,\hat\ed &=& 
   2\tanh\rho\,
   \left(\frac{1}{2}\hat{\pi}^{\eta\eta} - \hat{\cal P}_0\right),
\label{eq-gubser-evolution-e} 
\\
   \tau _{\pi}\partial_\rho\hat \pi^{\eta\eta}+\hat \pi^{\eta\eta} &=& 2\tanh\rho\left(\frac{2}{3}\hat{\eta}-\delta_{\pi\pi}\hat \pi^{\eta\eta}+\frac{1}{6}\tau_{\pi\pi}\hat \pi^{\eta\eta}\right),  
\label{eq-gubser-evolution-pi}
\\
   \partial_\rho\hat \n + 2\tanh\rho\,\hat{\n}&=&0\;,
\label{eq-gubser-evolution-n}
\\
   \tau _{\V}\partial_\rho\hat \V^\eta +\hat \V^\eta &=& -2\tanh\rho\left(\delta_{\V\V} - \frac{1}{3}\lambda_{\V\V}    
             \right) \hat \V^\eta\;.
\label{eq-gubser-evolution-neta}
\end{eqnarray}
(Note that $\hat\theta=2\tanh\rho$ is the scalar expansion rate for Gubser flow.) For the transport coefficients we use the same parametrization as described before in Fig.~\ref{F5} (see also the caption of Fig.~\ref{F6}). The transformation rules for the baryon density and diffusion current are
\begin{eqnarray}
   \n(\tau, r) &=& \hat\n(\rho(\tau, r))/\tau^3 \;,
\label{eq-gubser-n}
\\
   \V^\eta(\tau, r) &=& \hat\V^\eta(\rho(\tau, r))/\tau^4\;.
\label{eq-gubser-nmu}
\end{eqnarray}

Given initial conditions $\hat\ed(\rho_0)$, $\hat\pi^{\eta\eta}(\rho_0)$, $\hat\n(\rho_0)$, and $\hat\V^\eta(\rho_0)$ at, say, $\rho_0 = 0$, obtaining $\hat\peq(\rho_0)$ and $\hat T(\rho_0)$ with the help of EOS1, Eqs.~(\ref{eq-gubser-evolution-e}-\ref{eq-gubser-evolution-neta}) can be solved with {\sc Mathematica} in Gubser coordinates and then translated into Milne coordinates $(\tau,x,y,\eta_s=0)$ using the transformation rules (\ref{eq-gubser-umu}-\ref{eq-gubser-energy}) and (\ref{eq-gubser-n}-\ref{eq-gubser-nmu}). This semi-analytic solution can then be used to obtain initial conditions for \code\ on an initial proper time hypersurface $\tau_0$ which are then further evolved with the (3+1)-dimensional \code\ code. Specifically, \code\ requires initial data at $\tau_0$ for 
    $$\ed, \peq, T, u^x, u^y, \pi^{xx}, \pi^{yy}, \pi^{xy}, \pi^{\tau\tau}, \pi^{\tau x}, \pi^{\tau y}, \pi^{\eta\eta}, \n, n^\eta$$
on the computational $(x, y)$ grid (due to longitudinal boost-invariance these are only required at $\eta_s=0$). All remaining hydrodynamic components are either zero or can be obtained from the above by symmetry. For example, all shear stress components $\pi^{\mu\nu}(\tau_0, x, y, \eta_s=0)$ for \code\ can be obtained from the semi-analytic solution $\hat\pi^{\eta\eta}(\rho)$ by using tracelessness and azimuthal symmetry,
\begin{align}
    \hat\pi_{\theta\theta}(\rho)&=-\frac{1}{2}\cosh^2\rho\;\hat\pi_{\eta\eta}(\rho)\;, \quad
    \hat\pi_{\phi\phi}(\rho)=-\frac{1}{2}\cosh^2\rho\sin^2\theta\;\hat\pi_{\eta\eta}(\rho)
\end{align}
(with all other Gubser components being zero by symmetry), followed by 
\begin{equation}
    \pi_{\mu\nu}(\rho(\tau, r(x, y))) =\frac{1}{\tau^2}\frac{\partial \hat x^\alpha}{\partial x^\mu}\frac{\partial \hat x^\beta}{\partial x^\nu}\hat\pi_{\alpha\beta}(\rho)\;.
\end{equation}
This gives, for example, at $(\tau{=}\tau_0,\eta_s{=}0)$
\begin{equation}
    \pi^{\tau\tau}\Bigl(\rho(\tau_0, r(x, y))\Bigr) = -\frac{q^2\sin^2\theta}{2\tau_0^2}\,\hat\pi_{\eta\eta}(\rho)\;,\label{eq-gubseric}
\end{equation}
where the value of $\rho$ depends on the transverse grid point $(x,y)$.\footnote{%
    Obviously, whenever the transport coefficients are changed in \code, the semi-analytic solution must be recomputed accordingly for comparison, also because the full exact solution (not just its initial conditions in Gubser coordinates) is required to obtain initial conditions for the \code\ code in Milne coordinates.}

%

In Fig.~\ref{F6} we compare the \code\ output with the semi-analytic Gubser solution for the default setup described in the figure caption. One observes excellent agreement. Owing to the non-trivial transverse expansion of Gubser flow it allows to test additional source terms in the \code\ evolution equations when compared with the Bjorken flow test. We can also use it to study the performance of the root-finding algorithm in \code. 

As discussed in Sec.~\ref{sec-root}, two different methods are used in complementary ranges of the flow velocity separated by the critical value $v = 0.563624$ or, equivalently, $u^\tau = 1.21061$. Fig.~\ref{F6}(e) shows that the root-finding algorithm works equally well on both sides of the critical value of $u^\tau$. We also checked the precision and relative speed of convergence of the Newton-Raphson and modified iteration schemes described in Sec.~\ref{sec-root}. For both methods, the maximum number of iterations is set to 100 (which is never reached), and the root finding stops when the relative error $|v_{i+1}-v_{i}|/v_{i+1} < 10^{-6}$ or $|u^\tau_{i+1}-u^\tau_{i}|/u^\tau_{i+1} < 10^{-4}$. For the conformal EoS used here we found both methods to converge about equally well to the same result (within the specified uncertainty), with the modified iteration scheme being about 15\% faster than the Newton-Raphson method.\footnote{%
    During the early evolution stages the Newton-Raphson method converges somewhat faster but at later times the modified interaction scheme is found to be more efficient.} 


%
\subsection{Comparison to other codes}
\label{sec4.4}
%

Owing to their shared assumption of longitudinal boost-invariance, none of the tests described in the preceding subsections addresses the performance of the \code\ code in describing the expansion along the rapidity direction. To remedy this we have compared \code\ output for a transversally homogeneous system undergoing arbitrary longitudinal expansion without transverse expansion with the results from an independent (1+1)D hydrodynamic code developed by Monnai \cite{PhysRevC.86.014908}. To be able to study baryon number transport, EOS4 is used in both codes. A similar comparison was also made in Ref. \cite{Denicol:2018wdp} to test the performance of {\sc MUSIC}. Rather than directly using Monnai's code \cite{PhysRevC.86.014908} we compare our \code\ results with those reported in the comparison \cite{Denicol:2018wdp} with {\sc MUSIC}. In this sense, the following test is also a code comparison with {\sc MUSIC}. Qualitatively consistent results from an earlier \code\ study were already reported in \cite{Du:2018mpf}.

\begin{figure*}[!htbp]
\centering
\includegraphics[width= 0.9\textwidth]{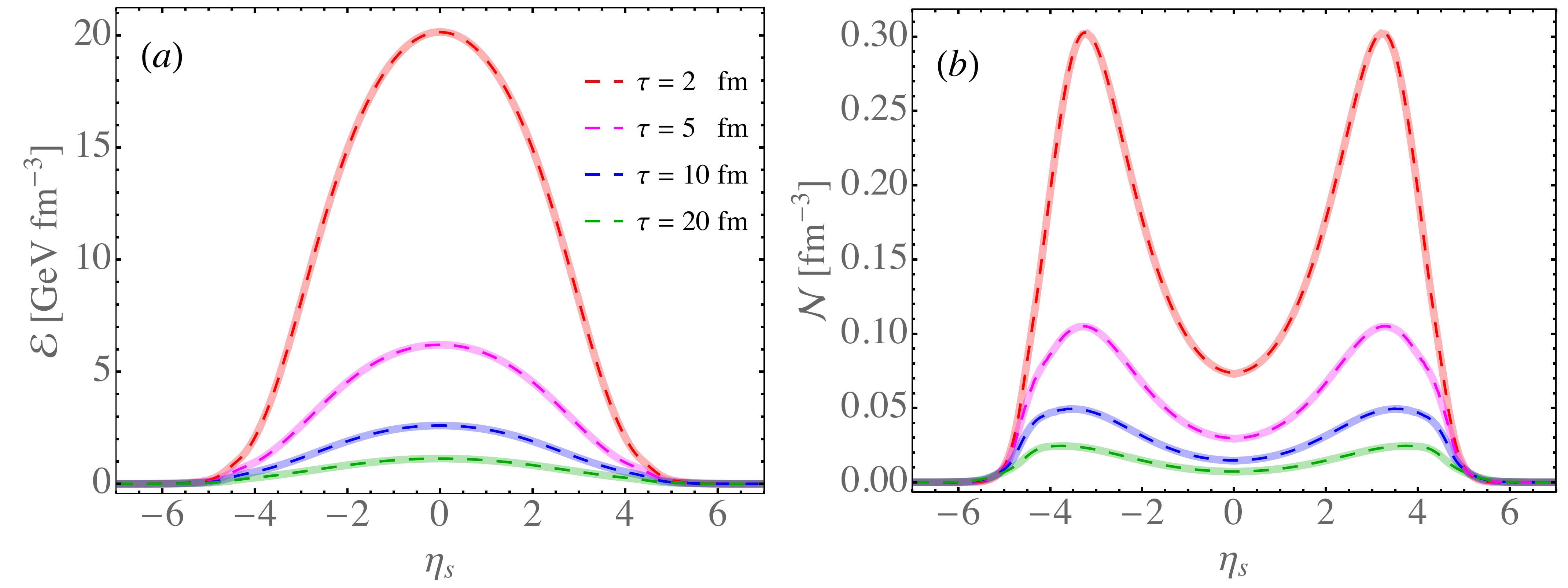}
\caption{Comparison between the numerical results from \code\ (broken lines) and the {\sc MUSIC} simulations \cite{Denicol:2018wdp} of the (1+1)D setup described in Ref.~\cite{PhysRevC.86.014908} (continuous lines): (a) energy density; (b) net baryon density. The simulation starts at $\tau_0 = 1$\,fm/$c$ and covers the space-time rapidity range $\eta_s\in[-6.94, 6.94]$, with grid spacing $\Delta\eta_s = 0.02$.
\label{F7}
}
\end{figure*}

We here focus on the baryon number evolution in the longitudinal direction, with non-vanishing longitudinal gradients of $\mu/T$, by setting bulk and shear stresses to zero.\footnote{%
    In principle, since shear stress is known to affect the evolution of the energy density it might be expected to also change the baryon number flow; this is an interesting physics question which we postpone for a separate study.}
Specifically, we check how \code\ handles the evolution equation (\ref{relEqs_nmu}) for baryon number diffusion,
\begin{equation}
    d \V^\mu = \frac{\kappa_n}{\tau_\V} \nabla^{\mu} \left(\frac{\mu}{T}\right) - \frac{\V^{\mu}}{\tau_{\V}} - n^\nu u^\mu D u_\nu 
    - u^\alpha \Gamma^\mu_{\alpha\beta} n^\beta\,,
\end{equation}
together with the net baryon conservation law (\ref{relEqs_n}). For the baryon transport coefficients we choose $\kappa_n = 0.2\,\n/\mu$ and $\tau_n = 0.2/T$. 

In Fig. \ref{F7} we show a comparison of the distributions in space-time rapidity $\eta_s$ of energy density $\ed$ and net baryon density $\n$ at four different times. Broken (continuous) lines show the
results from \code\ ({\sc MUSIC} \cite{Denicol:2018wdp}); the agreement between these two codes is very good. In addition to testing the longitudinal dynamics this comparison also shows that the root-finding algorithm works correctly with a realistic EoS\, $\peq(\ed,\n)$ that depends on both energy and net baryon density. 

\subsection{Tests summary}
\label{sec4.5}
We briefly summarize which parts of the \code\ code were tested with the test protocols described in this section. As described in Sec.~\ref{numericalscheme}, the same RK-KT algorithm is applied for solving the equations of motion for all hydrodynamical variables propagating in the system, and the same root-finding algorithm is used for all equations of state, whether they depend on baryon density or not. The observed good performance in propagating all hydrodynamical variables indicates the efficiency of the RK-KT algorithm, and the proper evolution of baryon density and baryon diffusion validates the root-finding method in systems with non-zero baryon currents. 

The Riemann problem in ideal hydrodynamics with EOS1 (Sec.~\ref{sec4.1}) shows the ability of the RK-KT algorithm in \code\ to capture shocks and contact discontinuities. The (0+1)D Bjorken expansion with EOS2 (Sec.~\ref{sec4.2}) tests the programming of the equations of motion, especially for the shear and bulk components, including the non-trivial parametrization of the transport coefficients, and the root-finding algorithm with non-zero but decoupled baryon density and diffusion current. The Gubser flow test with EOS1 (Sec.~\ref{gubsertest}) provides extra validation in situations with strong transverse expansion featuring large temporal and transverse gradients. The comparison  with {\sc MUSIC} and Monnai's (1+1)D code (Sec.~\ref{sec4.4}) at finite baryon density with the realistic EOS4 validates the longitudinal dynamics of density and baryon diffusion without the simplification of longitudinal boost-invariance, as well as the root-finding algorithm with non-zero baryon density and baryon diffusion, including the bilinear interpolation of the EoS tables. The figures shown in this section demonstrate that \code\ passes all these tests without struggle.

\section{Baryon diffusion in an expanding QGP}
\label{sec5}

In this section we illustrate the evolution of energy and baryon number in an expanding QGP with realistic ``bumpy'' initial conditions, by visualizing the evolution of the corresponding densities in the transverse plane at $\eta_s=0$.\footnote{%
    For a similar earlier study with smooth, ensemble-averaged initial conditions see \cite{Denicol:2018wdp}.}
This generalizes many similar visualizations made in the past for systems without conserved charges. Since the physics of initial-state fluctuations along the longitudinal direction is still not very well explored (for a few examples see, Refs.~\cite{Denicol:2015nhu, PhysRevC.86.014908, Denicol:2018wdp}), we here use smooth longitudinal initial conditions and refrain from showing the (mostly uninteresting) evolution along the beam direction.

Following Refs.~\cite{Shen:2017ruz, Du:2018mpf, Denicol:2018wdp} we use a 3-dimensional initial condition at non-zero baryon density which extends a transverse profile obtained from the Monte Carlo (MC) Glauber model \cite{Miller:2007ri} into the longitudinal direction with the following prescription:
\begin{eqnarray}
    \ed(\tau_0,x,y,\eta_s) &=& \frac{\ed_0}{\tau_0}\,
    \Bigl[T_A(x,y)\ed_A(\eta_s)+T_B(x,y)\ed_B(\eta_s)\Bigr]\;,
\\
    \n(\tau_0,x,y,\eta_s) &=& \frac{1}{\tau_0}\,
    \Bigl[T_A(x,y)\n_A(\eta_s)+T_B(x,y)\n_B(\eta_s)\Bigr]\;,
\end{eqnarray}
where $T_{A/B}(x,y)$ are the transverse profiles of the right- and left-moving nuclei from the MC-Glauber model and $\ed_{A/B}(\eta_s)$, $\n_{A/B}(\eta_s)$ are the corresponding longitudinal profiles for the energy and net baryon density, respectively \cite{Shen:2017ruz, Du:2018mpf, Denicol:2018wdp}. $\ed_0$ is a normalization factor which can be tuned to reproduce the final multiplicity while $\n(\tau_0,x,y,\eta_s)$ is normalized to the total number of participant baryons \cite{Denicol:2018wdp}.

\begin{figure}[!htpb]
    \centering
    \includegraphics[width=0.9\textwidth]{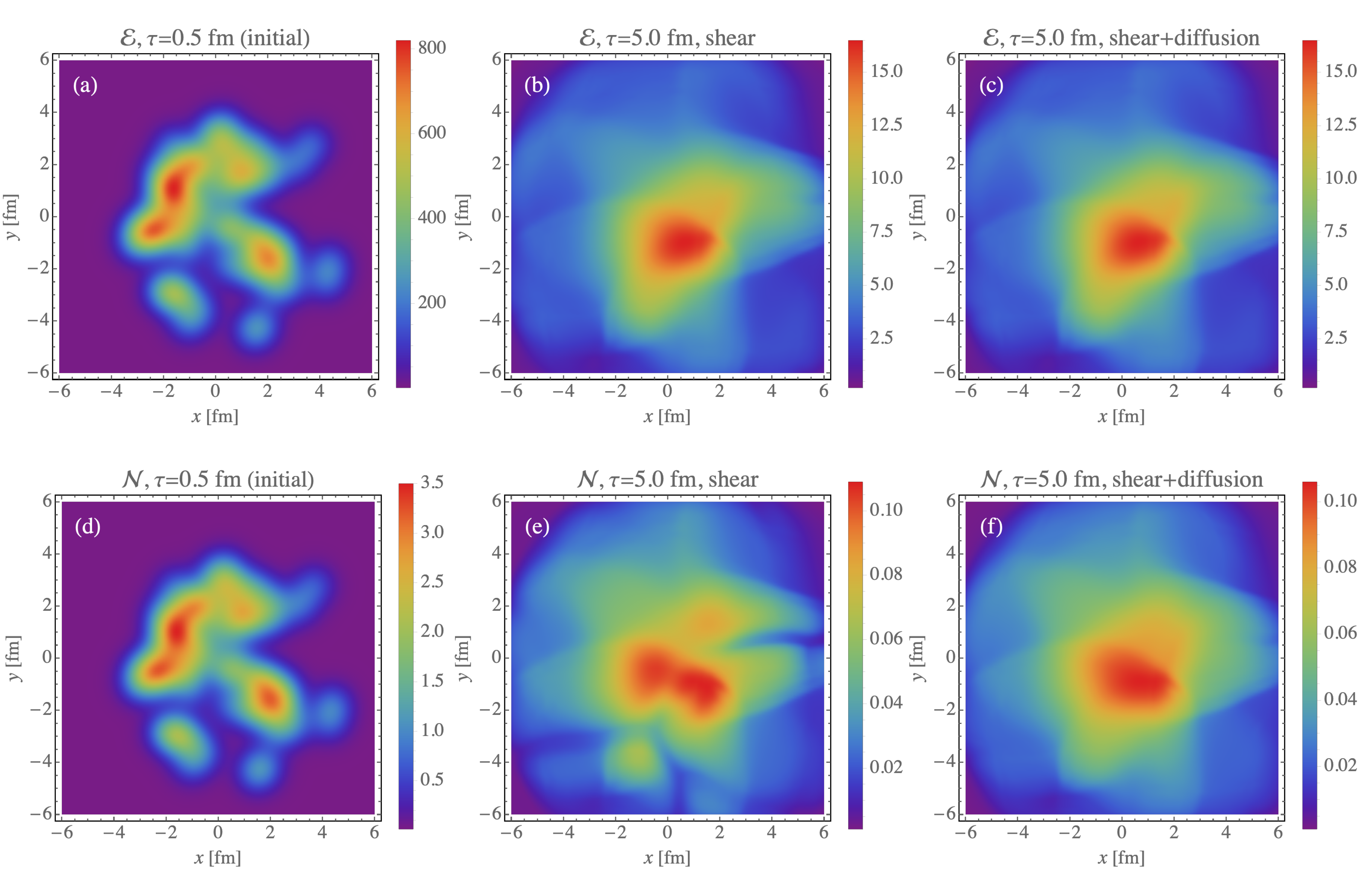}
    \caption{(Color online) Evolution of an expanding QGP with non-zero net baryon density with a bumpy initial condition, for a Cu+Cu collision at $b=4$\,fm. {\sl Top row:} Energy density $\ed$ in fm$^{-4}$ at initial time $\tau_0=0.5$\,fm/$c$ (a) and at time $\tau=5$\,fm/$c$ with kinematic shear viscosity $\bar\eta = 0.2$ and zero (b) or non-zero (c) baryon diffusion. {\sl Bottom row:} Same as top row, but for the net baryon density $\n$ in fm$^{-3}$. EOS4 is used for the equation of state, and $C_B = 0.4$ for evaluating the baryon diffusion coefficient $\kappa_n$ in Eq.~(\ref{kappa}). The bulk viscous pressure is set to zero.}
    \label{fig:visualization}
\end{figure}

In Fig. \ref{fig:visualization}, we show the transverse distributions at $\eta_s=0$ of the energy density (upper panels) and net baryon density (lower panels), at the hydrodynamic starting time $\tau_0 =0.5$\,fm/$c$ (left column) and later at $\tau=5$\,fm/$c$, for evolution with only shear stress turned on (middle column) and both shear stress and baryon diffusion turned on (right column). By comparing the middle and right columns we note that baryon diffusion leaves no pronounced signatures in the evolution of the energy density but smoothes out gradients in baryon density. The authors of Refs.~\cite{Shen:2017ruz,Du:2018mpf,Denicol:2018wdp} came to a similar conclusion for the evolution in the longitudinal direction.

\section{Conclusions and outlook}

In this work we have described the structure and performance of a new code called \code\ describing the (3+1)-dimensional space-time evolution of hot and dense matter created in relativistic heavy-ion collisions using second-order dissipative relativistic fluid dynamics. \code\ differs from most other publicly available algorithms by evolving, together with energy and momentum, a conserved current related to net baryon number, including its dissipative diffusion current, and it evolves the medium with an equation of state that depends on both the energy and net baryon densities. The generalization of the \code\ algorithm to the simultaneous propagation of multiple conserved charge currents \cite{Greif:2017byw} should be a straightforward task for the near future. A dynamical initialization module for \code\ that accounts for the non-zero temporal extension of the energy deposition process in heavy-ion collisions at collision energies probed in the Beam Energy Scan (BES) campaign at the Relativistic Heavy-Ion Collider, along the lines discussed in Refs.~\cite{Shen:2017bsr, Du:2018mpf}, is under construction and will soon be added to the \code\ package. The code can also be plugged in the JETSCAPE framework \cite{Putschke:2019yrg} as a hydrodynamic evolution module.

After briefly describing the physics to be addressed with \code\ simulations, we presented in detail the equations \code\ solves, the transport coefficients and equations of state it uses, and the numerical schemes it employs, including the root-finding algorithm for finding the flow velocity field at each time step and the regulation schemes used to regulate overly large dissipative flows caused by numerical or anomalously large physical fluctuations. The performance of the code was verified with high precision using a series of standard tests involving comparison with analytically or semi-analytically known solutions for problems of reduced dimensionality, characterized by additional symmetries that are usually not respected in real life situations but can be exploited for code verification. The code is distributed together with this suite of verification test protocols, thus enabling the user to check the continued accuracy of the code after changing or generalizing some of its parts. Finally, we presented a simple example illustrating the practical effects of baryon number diffusion on the evolution of energy and net baryon density for a collision between copper (Cu) nuclei, demonstrating the smoothing effects of baryon diffusion on large gradients of the net baryon density in the transverse plane.    


\section*{Acknowledgements} 
The authors thank Dennis Bazow, Derek Everett, Mike McNelis, Long-Gang Pang, Dirk Rischke, Chun Shen, Leonardo Tinti and Gojko Vujanovic for fruitful discussions. This work was supported in part by the U.S. Department of Energy (DOE), Office of Science, Office for Nuclear Physics under Award No. \rm{DE-SC0004286} and within the framework of the BEST Collaboration, and in part by the National Science Foundation (NSF) within the framework of the JETSCAPE Collaboration under Award No. \rm{ACI-1550223}. LD appreciates the kind hospitality of the European Centre for Theoretical Studies in Nuclear Physics and Related Areas (ECT*) and the MIT Center for Theoretical Physics during the completion of this work. UH's stay at the Institut f\"{u}r Theoretische Physik of the J. W. Goethe-Universit\"{a}t was supported by the Alexander von Humboldt Foundation through a Research Prize. Computing resources were generously provided by the Ohio Supercomputer Center \cite{OhioSupercomputerCenter1987} and by the Open Science Grid \cite{Pordes:2007zzb, Sfiligoi:2010zz}, which is supported by the National Science Foundation, award 1148698, and the U.S. Department of Energy's Office of Science.


\newpage
\begin{appendices}
\section{Conservative form of the evolution equations}
\label{appa}
In this Appendix, we recast Eqs.~(\ref{relEqs_Tmut}-\ref{relEqs_n}) and  (\ref{relEqs_Pi}-\ref{relEqs_pi}) in the conservative form (\ref{numericalEquations}). The procedure follows Refs.~\cite{Bazow:2016yra, Molnar:2009tx}, adding here the equations for baryon evolution. The equations reproduced here are written in a form that facilitates direct comparison with the \code\ code.

We start with the conservation laws in Eqs.~(\ref{relEqs_Tmut}-\ref{relEqs_n}). With the scaled flow velocities $v^i=u^i/u^\tau$ $(i=x,y,\eta_s)$, we can write down the following constituent relations for the components of $T^{\mu\nu}$ and $N^\mu$:
\begin{align}
T^{\tau\tau}&=(\ed+\p)u^{\tau}u^{\tau}-\p+\pi^{\tau\tau}\;,
\label{Ttt} \\
T^{\tau i}&=(\ed+\p)u^{\tau}u^{i}+\pi^{\tau i} =v^{i}T^{\tau\tau}+\p v^{i}-v^{i}\pi^{\tau\tau}+\pi^{\tau i}\;,
\label{Tti} \\
T^{ij}&=(\ed+\p)u^{i}u^{j}-\p g^{ij}+\pi^{ij} =v^{i}T^{\tau j}-\p g^{ij}-v^{i}\pi^{\tau i}+\pi^{ij}\;,
\label{Tij} \\
N^\tau &= \n u^\tau + n^\tau\;, \\
N^i &=\n u^i + n^i = v^i N^\tau - v^i n^\tau + n^i\;.
\end{align}
Inserting these into the conservation laws one obtains
\begin{align}
\partial _{\tau }T^{\tau \tau }+\partial _{x}(v^{x}T^{\tau \tau})+\partial_{y}(v^{y}T^{\tau \tau })+\partial _{\eta }(v^{\eta}T^{\tau
\tau }) &= I^\tau_2 + I^\tau_x +I^\tau_y +I^\tau_\eta\,, 
\\%
\partial _{\tau }T^{\tau x}+\partial _{x}(v^{x}T^{\tau x})+\partial
_{y}(v^{y}T^{\tau x})+\partial _{\eta }(v^{\eta}T^{\tau x}) &= I^x_2 + I^x_x +I^x_y +I^x_\eta\,,
\\%
\partial _{\tau }T^{\tau y}+\partial _{x}(v^{x}T^{\tau y})+\partial
_{y}(v^{y}T^{\tau y})+\partial _{\eta }(v^{\eta}T^{\tau y}) &= I^y_2 + I^y_x +I^y_y +I^y_\eta \,, 
\\%
\partial _{\tau }T^{\tau \eta }+\partial _{x}(v^{x}T^{\tau \eta
})+\partial _{y}(v^{y}T^{\tau \eta })+\partial _{\eta }(v^{\eta}T^{\tau
\eta }) &= I^\eta_2 + I^\eta_x +I^\eta_y +I^\eta_\eta \,, \label{dtT03}
\\%
\partial _{\tau }N^{\tau}+\partial _{x}(v^{x}N^{\tau})+\partial _{y}(v^{y}N^{\tau})+\partial _{\eta }(v^{\eta}N^{\tau})&= J^\tau_2 + J^\tau_x +J^\tau_y +J^\tau_\eta\;, \label{baryonconservation}
\end{align}
with the following source terms for $T^{\tau\tau }$:
\begin{align}
I^\tau_2=&-\frac{1}{\tau }\left( T^{\tau \tau }+\tau ^{2}T^{\eta \eta}\right)-\left(\peq+\Pi-\pi ^{\tau \tau }\right)\partial _{i}v^i-v^i\partial_i \peq\,, \\
I^\tau_x= &-v^{x}\partial _{x}\left(\Pi-\pi ^{\tau \tau }\right)-\partial _{x}\pi ^{\tau x}\,, \\
I^\tau_y= &-v^{y}\partial _{y}\left(\Pi-\pi ^{\tau \tau }\right)-\partial _{y}\pi ^{\tau y}\,, \\
I^\tau_\eta= &-v^{\eta}\partial _{\eta}\left(\Pi-\pi ^{\tau \tau }\right)-\partial _{\eta}\pi ^{\tau \eta}\,; 
\end{align}
for $T^{\tau x}$:
\begin{align}
I^x_2=&-\frac{1}{\tau }T^{\tau x}-\partial_{x}\peq +\pi ^{\tau x}\partial _{i}v^i\,, \\
I^x_x= &-\partial _{x}\left( \Pi+\pi ^{xx}\right)+v^{x}\partial _{x}\pi ^{\tau x}\,,\\
I^x_y =  & -\partial_{y}\pi ^{xy}+v^{y}\partial_{y}\pi ^{\tau x}\,, \\
I^x_\eta=  &-\partial_{\eta}\pi ^{x\eta}+v^{\eta}\partial_{\eta}\pi ^{\tau x}\,;
\end{align}
for $T^{\tau y}$:
\begin{align}
I^y_2=&-\frac{1}{\tau }T^{\tau y}-\partial_{y}\peq +\pi ^{\tau y}\partial _{i}v^i\,,\\
I^y_x=&-\partial _{x}\pi ^{xy}+v^{x}\partial _{x}\pi ^{\tau y}\,,\\
I^y_y=&-\partial_{y}\left(\Pi+\pi ^{yy}\right)+v^{y}\partial_{y}\pi ^{\tau y}\,, \\
I^y_\eta=&-\partial_{\eta}\pi ^{y\eta}+v^{\eta}\partial_{\eta}\pi ^{\tau y}\,; 
\end{align}
for $T^{\tau \eta}$:
\begin{align}
I^\eta_2=&-\frac{3}{\tau}T^{\tau\eta}
          -\frac{\partial_{\eta}\peq}{\tau^{2}}
          +\pi^{\tau\eta}\partial_{i}v^i\,,
\\
I^\eta_x=&-\partial_{x}\pi^{x\eta}+v^{x}\partial_{x}\pi^{\tau\eta}\,,
\\
I^\eta_y=&-\partial_{y}\pi ^{y\eta}+v^{y}\partial_{y}\pi^{\tau\eta}\,,
\\
I^\eta_\eta=&-\partial_{\eta }\left(\Pi/\tau ^{2}+\pi^{\eta\eta}\right)
+ v^{\eta}\partial_{\eta }\pi^{\tau \eta} \,;
\end{align}
and for $N^\tau$:
\begin{align}
J^\tau_2=&-\frac{1}{\tau } N^{\tau}+n^\tau\partial _{i}v^i \,,
\\
J^\tau_x=&-\partial _{x}n ^{x} +v^{x}\partial _{x}n ^{\tau}\,,
\\
J^\tau_y=& -\partial _{y}n^{y} +v^{y}\partial _{y}n^{\tau}\,,
\\
J^\tau_\eta=& -\partial_{\eta }n^{\eta} 
           +v^{\eta} \partial_{\eta}n^{\tau}\;. 
\end{align}
Considering $d \equiv u^\mu \partial_\mu$, the relaxation equations (\ref{relEqs_Pi}-\ref{relEqs_pi}) for the dissipative flows can be written as
\begin{eqnarray}
    \partial_{\tau}\Pi+\partial_{x}(v^{x}\Pi)+\partial_{y}(v^{y}\Pi)+\partial_{\eta}(v^{\eta}\Pi)&=&S^\Pi_2 \;,
\label{relEqs_Pia}\\
    \partial_{\tau}\V^\mu+\partial_{x}(v^{x}\V^\mu)+\partial_{y}(v^{y}\V^\mu)+\partial_{\eta}(v^{\eta}\V^\mu) &=&S^\V_2
\label{relEqs_na}\;,\\
    \partial_{\tau}\pi^{\mu\nu} +\partial_{x}(v^{x}\pi^{\mu\nu})+\partial_{y}(v^{y}\pi^{\mu\nu})+\partial_{\eta}(v^{\eta}\pi^{\mu\nu}) &=&S^\pi_2 \;,
\label{relEqs_pia}
\end{eqnarray}
where we used $\partial_{i}v^{i} \equiv \partial_{x}v^{x} + \partial_{y}v^{y} + \partial_{\eta}v^{\eta}$. The source terms in Eqs.~(\ref{relEqs_Pia}-\ref{relEqs_pia}) are given by 
\begin{eqnarray}
    S^\Pi_2&=&\frac{1}{u^\tau}\left(-\frac{\zeta}{\tau_\Pi}\theta-\frac{\Pi}{\tau_\Pi}-I_\Pi\right)+\Pi\partial_i v^i
    \label{relEqs_SPia}\;,\\
    S^\V_2&=&\frac{1}{u^\tau}\left(\frac{\kappa_n}{\tau_\V} \nabla^{\mu }\left(\frac{\mu}{T}\right)-\frac{\V^{\mu}}{\tau _{\V}}-I_\V^\mu-G_\V^\mu\right)+\V^\mu\partial_i v^i 
    \label{relEqs_Sna}\;,\\
    S^\pi_2&=&\frac{1}{u^\tau}\left(\frac{2\eta}{\tau_{\pi}}\sigma^{\mu\nu}-\frac{\pi^{\mu\nu}}{\tau_{\pi}}-I^{\mu\nu}_{\pi}-G^{\mu\nu}_{\pi}\right)+\pi^{\mu\nu}\partial_i v^i\;.
    \label{relEqs_Spia}
\end{eqnarray}
Here the $I$-terms and $G$-terms are defined in Eqs. (\ref{relEqs_Pi})-(\ref{relEqs_pi}).

\section{Explicit form}

In this Appendix we provide explicit forms of some expression occurring in the baryon evolution equations. In Milne coordinates the terms $G^\mu_n = u^\alpha \Gamma^\mu_{\alpha\beta} n^\beta$ in Eq.~(\ref{relEqs_nmu}) evaluate to 
\begin{align}
G^\tau_n &= \tau u^\eta n^\eta\,,\\
G^x_n &= 0\,,\\
G^y_n &= 0\,,\\
G^\tau_n &= (u^\tau n^\eta + u^\eta n^\tau)/\tau\,.
\end{align}
To calculate the LRF gradient of $\mu/T$ in Eq.~(\ref{relEqs_nmu}) numerically we use
\begin{equation}
 \nabla^{\mu}\left(\frac{\mu}{T}\right) \equiv \Delta^{\mu\nu} d_\nu\left(\frac{\mu}{T}\right) = (g^{\mu\nu}-u^\mu u^\nu)\partial_\nu\left(\frac{\mu}{T}\right)
\end{equation}
and work out the partial derivatives $\partial_\mu (\mu/T)$ in the computational frame numerically from the EoS tables. Finally, the last two terms in Eq.~(\ref{I1-I4-mu}) can be expressed as
\begin{align}
I_3^\mu& = n_{\nu }\omega ^{\nu \mu } = n^{\tau }\omega ^{\tau \mu } - n^x\omega ^{x \mu } - n^y\omega ^{y \mu } - \tau^2 n^{\eta }\omega ^{\eta \mu }\;,\\
I_4^\mu& = n_{\nu }\sigma ^{\nu \mu }= n^\tau\sigma^{\tau \mu } - n^x\sigma^{x \mu } - n^y\sigma^{y \mu }  - \tau^2 n^\eta \sigma^{\eta \mu }\;.
\end{align}
\end{appendices}

\bibliographystyle{elsarticle-num}
\bibliography{beshydro}
\end{document}